\input harvmac
\def\Str{{\rm Str}}
\overfullrule=0pt

%
\def\sqr#1#2{{\vbox{\hrule height.#2pt\hbox{\vrule width
.#2pt height#1pt \kern#1pt\vrule width.#2pt}\hrule height.#2pt}}}

\def\half{{\textstyle{1\over 2}}}

\Title{ \vbox{\baselineskip12pt
\hbox{hep-th/0411020}}}
{\vbox{\centerline{Spin Models }
\bigskip
\centerline{and Superconformal Yang-Mills Theory}}}
\smallskip
\centerline{Louise Dolan}
\smallskip
\centerline{\it Department of Physics}
\centerline{\it
University of North Carolina, Chapel Hill, NC 27599-3255}
\bigskip
\smallskip
\centerline{Chiara R. Nappi
}
\smallskip
\centerline{\it Department of Physics, Jadwin Hall}
\centerline{\it Princeton University, Princeton, NJ 08540}
\bigskip
\bigskip
\bigskip
\bigskip

\noindent 
We apply novel techniques in planar superconformal Yang-Mills theory
which stress the role of the Yangian algebra. 
We compute the first two Casimirs of the Yangian, which are identified
with the first two local abelian Hamiltonians with periodic boundary
conditions, and show that they annihilate the chiral primary states. 
We streamline the derivation of the R-matrix in a conventional spin model, 
and extend this computation to the gauge theory. 
We comment on higher-loop corrections and higher-loop integrability. 


\Date{}

\nref\Minahan{J.~A.~Minahan and K.~Zarembo, ``The Bethe-Ansatz for
N = 4 Super Yang-Mills,'' JHEP {\bf 0303}, 013 (2003)
hep-th/0212208.}

\nref\Beisertone{N. Beisert, ``The Complete One-loop Dilatation
Operator of N = 4 super Yang-Mills Theory'',
Nucl.\ Phys.\ B {\bf 676}, 3 (2004), hep-th/0307015.}

\nref\Beiserttwo{N.~Beisert and M.~Staudacher, ``The N = 4 SYM
Integrable Super Spin Chain,'' 
Nucl.\ Phys.\ B {\bf 670}, 439 (2003), hep-th/0307042.}

\nref\Beisertthree{N.~Beisert, J.~A.~Minahan, M.~Staudacher and
K.~Zarembo, ``Stringing Spins and Spinning Strings,''
JHEP {\bf 0309}, 010 (2003), hep-th/0306139.}

\nref\Beisertfour{N.~Beisert, C.~Kristjansen and M.~Staudacher,
``The Dilatation Operator of N = 4 Super Yang-Mills Theory,''
Nucl.\ Phys.\ {\bf B664}, 131 (2003), hep-th/0303060.}

\nref\Beisertsix{N.~Beisert,
``The Dilatation Operator of N = 4 Super Yang-Mills Theory and
Integrability,'' hep-th/0407277.}

\nref\Belitskyone{A.~Belitsky, A. ~Gorsky and G. ~Korchemsky,
``Gauge / String Duality for QCD Conformal Operators,'' 
Nucl.\ Phys.\ B {\bf 667}, 3 (2003), hep-th/0304028.}

\nref\Belitskytwo{A.~Belitsky, 
V. Braun, A. Gorsky and G. Korchemsky,
``Integrability in QCD and Beyond,'' hep-th/0407232.}

\nref\polyakovtwo{A.~M.~Polyakov,
``Confinement and Liberation,'' hep-th/0407209.}

\nref\polyakovthree{A.~M.~Polyakov,
``Conformal Fixed Points of Unidentified Gauge Theories,''
Mod.\ Phys.\ Lett.\ A {\bf 19}, 1649 (2004), hep-th/0405106.}

\nref\thooft{G. 't Hooft, ``A Planar Diagram Theory for Strong
Interactions,'' Nucl. Phys. {\bf B72}, 461 (1974).}

\nref\lipatov{L. N. Lipatov, ``High-Energy Asymptotics of Multicolor
QCD and Exactly Solvable Lattice Models,'' JETP Lett. {\bf 59}, 596 (1994),
hep-th/9311037.}

\nref\Korchemsky{G. Korchemsky and L. Faddeev,
`High-energy QCD as a Completely Integrable Model,''
Phys.\ Lett.\ B {\bf 342} 311 (1995), hep-th/9404173.}

\nref\bmn{D. Berenstein, J. Maldacena, and H. Nastase,
``Strings in Flat Space and PP Waves from $N=4$ Super Yang Mills,''
JHEP {\bf 204}, 013 (2002), hep-th/0202021.}

\nref\Kotikov{A.~V.~Kotikov, L.~N.~Lipatov and V.~N.~Velizhanin,
``Anomalous Dimensions of Wilson Operators in N = 4 SYM theory,''
Phys.\ Lett.\ B {\bf 557} 114 (2003), hep-ph/0301021.}

\nref\Osborn{F.~A.~Dolan and H.~Osborn,
``Superconformal Symmetry, Correlation Functions and the Operator Product
Expansion,''
Nucl.\ Phys.\ B {\bf 629}, 3 (2002), hep-th/0112251.}

\nref\Belitskya{A.~V.~Belitsky,
``Fine Structure of Spectrum of Twist-Three Operators in {QCD},''
Phys.\ Lett.\ B {\bf 453} 59 (1999), hep-ph/9902361;
``Integrability and WKB Solution of Twist-Three Evolution Equations,''
Nucl.\ Phys.\ B {\bf 558} 259 (1999), hep-ph/9903512;
`Renormalization of Twist-Three Operators and Integrable Lattice Models,''
Nucl.\ Phys.\ B {\bf 574} 407 (2000), hep-ph/9907420.}

\nref\Braun{V.~M.~Braun, S.~E.~Derkachov and A.~N.~Manashov,
``Integrability of Three-Particle Evolution Equations in {QCD},''
Phys.\ Rev.\ Lett.\  {\bf 81}  2020 (1998), hep-ph/9805225.}

\nref\Manashov{V.~M.~Braun, S.~E.~Derkachov, G.~P.~Korchemsky
and A.~N.~Manashov,
``Baryon Distribution Amplitudes in {QCD},''
Nucl.\ Phys.\ B {\bf 553} 355 (1999), hep-ph/9902375.}

\nref\Derkachov{S.~E.~Derkachov, G.~P.~Korchemsky and A.~N.~Manashov,
``Evolution Equations for Quark Gluon Distributions in Multi-color QCD
and Open Spin Chains,''
Nucl.\ Phys.\ B {\bf 566} 203 (2000), hep-ph/9909539.}

\nref\Derkachovtwo{A.~V.~Belitsky, S.~E.~Derkachov, G.~P.~Korchemsky 
and A.~N.~Manashov,
``Superconformal Operators in N = 4 super-Yang-Mills Theory,''
Phys.\ Rev.\ D {\bf 70}, 045021 (2004), hep-th/0311104.}

\nref\Kazakov{V.~A.~Kazakov, A.~Marshakov, J.~A.~Minahan and K.~Zarembo,
``Classical / Quantum Integrability in AdS/CFT,''
JHEP {\bf 0405}, 024 (2004), hep-th/0402207.}

\nref\Roiban{R.~Roiban and A.~Volovich,
``Yang-Mills Correlation Functions from Integrable Spin Chains,''
hep-th/0407140.}

\nref\kazzar{V.~A.~Kazakov and K.~Zarembo,
``Classical/Quantum Integrability in the Non-Compact Sector of AdS/CFT,''
hep-th/0410105.}

\nref\DNW{L. Dolan, C. Nappi, and E. Witten,
``A Relation Between Approaches to Integrability in Superconformal
Yang Mills Theory'', JHEP{\bf 0310}, 017 (2003),
hep-th/0308089.}   

\nref\cinn{L. Dolan, C. Nappi, and E. Witten,
``Yangian Symmetry in $D=4$ Superconformal Yang-Mills Theory'',
QTS3 Conference Proceedings, University of Cincinnati, September 2003,
hep-th/0401243.}

\nref\DRone{V. Drinfel'd, ``Hopf Algebras and the Quantum Yang-Baxter
Equation'', Sov. Math. Dokl. {\bf 32} 254 (1985).}

\nref\DRtwo{V. Drinfel'd, ``A New Realization of Yangians and
Quantized Affine Algebras'',  Sov. Math. Dokl. {\bf 36} 212 (1988).}

\nref\DRthree{V. Drinfel'd, ``Quantum Groups'', Proceedings of the
International Congress of Math, Berkeley, California, 1986, pp 798-820.}

\nref\Bernardtwo{D. Bernard,
``An Introduction to Yangian Symmetries,''
\ Int. \ J.\ Mod.\ Phys.  {\bf B7}, 3517 (1993)
hep-th/9211133.}

\nref\Bernardthree{D. Bernard and O. Babelon, ``Dressing Symmetries'',
Commun.\ Math.\ Phys.\  {\bf 149}, 279 (1992),
hep-th/9111036.}

\nref\Bernardone{D. Bernard,
``Hidden Yangian in 2D Massive Current Algebras,''
Commun.Math.Phys. {\bf 137}, 191 (1991).}

\nref\HH{F. Haldane, Z. Ha, J. Talstra, D. Bernard, and V. Pasquier,
``Yangian Symmetry of Integrable Quantum Chains with Long-Range
Interactions and a New Description of States in Conformal Field Theory'',
Phys. Rev. Lett. {\bf 69}, 2021 (1992).}

\nref\mackay{N. J.~MacKay,
``On the Classical Origins of Yangian Symmetry in Integrable Field Theory,''
Phys.\ Lett.\ B {\bf 281}, 90 (1992).}

\nref\mackaytwo{N. J.~MacKay, 
``Introduction to Yangian Symmetry in Integrable Field Theory'',
hep-th/0409183.}

\nref\wadia{G. Mandal, N. V. Suryanarayana, and S.R. Wadia,
``Aspects of Semiclassical Strings in AdS(5),'' Phys. Lett. {\bf B543},
81 (2002), hep-th/0206103.}

\nref\Metsaev{R.~R.~Metsaev and A.~A.~Tseytlin, ``Type IIB
Superstring Action in AdS(5) x S(5) Background,'' Nucl.\ Phys.\
{\bf B533}, 109 (1998), hep-th/9805028.}

\nref\Bena{I. Bena, J. Polchinski and R. Roiban, ``Hidden
Symmetries of the AdS(5) x $S^5$ Superstring,''
hep-th/0305116.}

\nref\KRS{P. Kulish, N. Reshetikhin, E. Sklyanin, ``Yang-Baxter Equation
and Representation Theory 1'',
Lett. Math. Phys. {\bf 5} 393 (1981).}

\nref\faddeev{ L. Takhtajan and L. Faddeev, {\it Hamiltonian
Methods in the Theory of Solitons}, Springer Series in Soviet
Mathematics, Springer Verlag, 1987; L. Faddeev, ``How Algebraic Bethe Ansatz
works for Integrable Models'', hep-th/9605187;  L. Faddeev,
``Algebraic Aspects of Bethe-Ansatz'', hep-th/9404013.}

\nref\faddeevfour{ L. Faddeev, ``Integrable Models in $(1+1)$-Dimensional
Quantum Field Theory'', in {\it Recent Advances in Field Theory and
Statistical Mechanics}, Proceedings of the Les Houches Summer School 1982,
North-Holland: Amsterdam 1984, pp. 561-608.}

\nref\Luschertwo{M. Luscher, ``Dynamical Charges in the Quantized,
Renormalized Massive Thirring Model'',
Nucl. Phys. {\bf B117}, 475 (1976).}

\nref\faddeevtwo{ V. Tarasov, L. Takhtajan, and L. Faddeev, 
``Local Hamiltonians for Integrable Quantum Models on a Lattice'',
Theor.\ Math.\ Phys.\  {\bf 57}, 1059 (1983).}

\nref\grabowskitwo{M. Grabowski and P. Mathieu, ```Structure of the 
Conservation Laws in Integrable Spin Chains with Short Range
Interactions,'' Annals Phys.\  {\bf 243}, 299 (1995)
hep-th/9411045.}

\nref\eks{E. Sklyanin, ``Boundary Conditions for Integrable Quantum
Systems'', J. Phys. A: Math Gen. {\bf 21} (1988) 2375.}

\nref\gm{M. Grabowski and P. Mathieu, ``The Structure of Conserved Charges
in Open Spin Chains'',  J. Phys. A: Math Gen. {\bf 29} (1996) 7635.}

\nref\Aharony{O.~Aharony, S.~S.~Gubser, J.~M.~Maldacena, H.~Ooguri and Y.~Oz,
``Large N Field Theories, String Theory and Gravity,''
Phys.\ Rept.\  {\bf 323}, 183 (2000)
hep-th/9905111.}

\nref\Beisertfive{N. Beisert, ``The $SU(2|3)$ Dynnamic Spin Chain'',
hep-th/0310252.}

\nref\Serban{D. Serban and M. Staudacher, ``Planar $N=4$ Gauge Theory and
the Inozemtsev Long Range Spin Chain'', JHEP {\bf 0406}, hep-th/0401067.}

\nref\Aru{G.~Arutyunov and M.~Staudacher,
``Two-loop Commuting Charges and the String / Gauge Duality,''
hep-th/0403077.}

\nref\Callanone{C. Callan,  T.~McLoughlin and I.~Swanson,
``Holography Beyond the Penrose Limit,''
Nucl.\ Phys.\ B {\bf 694}, 115 (2004)
hep-th/0404007.}

\nref\Callantwo{C.~G.~.~Callan, T.~McLoughlin and I.~Swanson,
``Higher Impurity AdS/CFT Correspondence in the Near-BMN Limit,''
hep-th/0405153.}

\nref\Swanson{I.~Swanson,
``On the Integrability of String Theory in AdS(5) x S**5,''hep-th/0405172.}


\nref\nm{N.~Mann and J.~Polchinski,
``Finite Density States in Integrable Conformal Field Theories,''
hep-th/0408162.}

\nref\frolov{N.~Beisert, S.~Frolov, M.~Staudacher and A.~A.~Tseytlin,
``Precision Spectroscopy of AdS/CFT,''
JHEP {\bf 0310}, 037 (2003), hep-th/0308117.}

\nref\tseytlin{A.~A.~Tseytlin, ``Semiclassical Strings and AdS/CFT,''
hep-th/0409296.}

\nref\Sadri{D.~Sadri and M.~M.~Sheikh-Jabbari,
``The Plane-Wave / Super Yang-Mills Duality,'' hep-th/0310119.}

\nref\Engquist{J.~Engquist, J.~A.~Minahan and K.~Zarembo,
``Yang-Mills Duals for Semiclassical Strings on AdS(5) x S**5,''
JHEP {\bf 0311}, 063 (2003), hep-th/0310188.}

\nref\Zarembo{M.~Lubcke and K.~Zarembo,
``Finite-size Corrections to Anomalous Dimensions in N = 4 SYM theory,''
JHEP {\bf 0405}, 049 (2004), hep-th/0405055.}

\nref\Arutyunov{G.~Arutyunov and M.~Staudacher,
``Matching Higher Conserved Charges for Strings and Spins,''
JHEP {\bf 0403}, 004 (2004), hep-th/0310182.}

\nref\Engtwo{J.~Engquist,
``Higher Conserved Charges and Integrability for Spinning Strings in 
AdS(5) x S**5,'' JHEP {\bf 0404}, 002 (2004), hep-th/0402092.}

\nref\Smedback{M.~Smedback, ``Pulsating Strings on AdS(5) x S**5,''
JHEP {\bf 0407}, 004 (2004), hep-th/0405102.}

\nref\Berktwo{N.~Berkovits,
``BRST Cohomology and Nonlocal Conserved Charges,''
hep-th/0409159.}

\nref\Polyakovone{A. M. Polyakov, ``Interaction of Goldstone
Particles in Two Dimensions. Applications to Ferromagnets and
Massive Yang-Mills Fields,'' Phys. Lett. {\bf B 59}, 79 (1975);
``Hidden Symmetry Of The Two-Dimensional Chiral Fields,''
Phys. Lett {\bf B72}, 224 (1977);
``String Representations and Hidden
Symmetries for Gauge Fields,'' Phys. Lett B82 (247) 1979;
``Gauge Fields as Rings of Glue,'' Nucl. Phys. {\bf B164}, 1971 (1980).}

\nref\LP{M. Luscher and K. Pohlmeyer, ``Scattering of Massless
Lumps and Nonlocal Charges in Two-dimensional Classical
Non-linear Sigma Model'',  Nucl. Phys. {\bf B137}, 46 (1978).}

\nref\Luscher{M. Luscher, ``Quantum Non-local Charges and Absence of Particle
Production in the Two-dimensional Non-linear $\sigma$ Model'',
\ Nucl.\  Phys. {\bf B135}, 1 (1978).}

\nref\Dolanone{L.
Dolan, ``Kac-Moody Algebra is Hidden Symmetry of Chiral Models,''
\ Phys.\ Rev. \ Lett. {\bf 47} 1371 (1981);
``Kac-Moody Symmetry Group of Real Self-dual
Yang-Mills,'' \ Phys. \ Lett. {\bf 113B}, 378 (1982);
``Kac-Moody Algebras and Exact
Solvability in Hadronic Physics'', \ Phys. \ Rep. {\bf 109}, 1
(1984).}

\nref\GO{P. .~Goddard and D.~I.~Olive,
``Kac-Moody And Virasoro Algebras: A Reprint Volume For Physicists,''
Adv.\ Ser.\ Math.\ Phys.\  {\bf 3} (1988) 1.}

\nref\schwarz{J.~H.~Schwarz,
``Classical Symmetries of Some Two-dimensional Models Coupled to Gravity,''
Nucl.\ Phys.\ B {\bf 454}, 427 (1995), hep-th/9506076.}

\nref\berkstu{B.~C.~Vallilo, ``Flat Currents in the Classical
${\rm AdS}_5 \times  S^5$ Pure Spinor Superstring,''
hep-th/0307018.}

\nref\berk{N.~Berkovits,
``Super-Poincare Covariant Quantization of the Superstring,''
JHEP {\bf 0004}, 018 (2000),
hep-th/0001035.}

\nref\Witten{E.~Witten,
``Perturbative Gauge Theory as a String Theory in Twistor Space,''
hep-th/0312171.}

\nref\Cach{F.~Cachazo, P.~Svrcek and E.~Witten,
``MHV Vertices and Tree Amplitudes in Gauge Theory,''
JHEP {\bf 0409}, 006 (2004), hep-th/0403047.}

\nref\Mason{L.~J.~Mason and N.~M.~J.~Woodhouse,
{\it Integrability, Self-Duality, and Twistor Theory},
Oxford, U.K. Clarendon (1996) 364 p. London Mathematical Society 
monographs, new series: 15.}

\newsec{Introduction}

Planar superconformal gauge theory in four dimensions has
yielded to a partial description in terms of integrable spin model 
structures. In the $N\rightarrow\infty$ limit, the ${\cal N}=4$ 
supersymmetric $SU(N)$ gauge theory has a dilatation operator whose
contribution at one-loop in $g^2 N$
can be identified with the Hamiltonian of an integrable spin chain
\refs{\Minahan - \kazzar}. Furthermore,
the ordinary $PSU(2,2| 4)$ symmetry generators of the gauge theory at
tree level, $g^2 N = 0$, can be identified with 
the total spin variables $J^A = \sum_{i=1} J^A_i$,
where $i$ labels the sites of the chain or lattice,
and $A$ runs over the dimension of the symmetry group. 
Spin models
were further exploited in  \refs{\DNW,\cinn } to uncover additional tree level
operators in the gauge theory,
$Q^A = f^A_{BC}\sum_{j<k} J^B_j J^C_k$, 
which generate the higher non-local charges
${\cal J}_n^A,\,\, n = 2,3,...,$ of a 
Yangian algebra \refs{ \DRone -\mackaytwo}.
Evidence for this larger Yangian symmetry 
had already been seen, at least at $g^2N=\infty$,
in the AdS/CFT dual theory described by
the classical Green-Schwarz superstring action for $AdS_5\times S^5$
\refs{\wadia-\Bena}. Another element of integrability in the gauge theory
is the R-matrix,
which was extrapolated from a subsector \refs{\Beiserttwo}. 
There are also one-loop higher local commuting Hamiltonians
which were computed in some special subsectors of the gauge theory
\refs{\Beisertfour}. 


With these identifications, it is compelling to ask how much
more of the gauge theory can be cast in integrable forms
that could eventually reveal
the exact spectrum and correlation functions. 

In this paper, we apply novel techniques in planar superconformal 
Yang Mills theory to analyze the integrable structure, and also show
how they streamline the calculations of conventional spin models
\refs{\KRS-\faddeev}.
We focus on the role of the Yangian in the gauge theory, 
and explain why it is useful even though it cannot be defined
for periodic spin chains. We use it to compute the R-matrix $R_{ij}(u)$,
and to define the local commuting Hamiltonians 
$H_\kappa$ with periodic boundary conditions, by identifying them as
Casimirs of the Yangian.

We compute the second Casimir explicitly in the
full gauge theory, and show
that it annihilates the chiral primary states. 
We conjecture that {\it all} the Casimirs of the
Yangian annihilate the chiral primaries. Since the supergravity states
in the AdS/CFT correspondence are
chiral primary states, this explains why 
one does not see the Yangian symmetry in the supergravity Lagrangian.

As an introduction to our methods, 
we first consider cases of more conventional 
quantum spin models and show how the Yangian can be used in those models to
find the R-matrix and Hamiltonians.
In sections 2 and 3, we treat the $SU(2)$ XXX-model in the spin
$1/2$ representation, and 
the $SL(2)$ chain in the spin $s$ representation.

In section 4 we compute both the R-matrix and the Hamiltonian directly
from the Yangian for
the $PSU(2,2|4)$ spin chain where the single-site spin variables are in  
the representation given by the one-particle states in free
${\cal N}=4$ super Yang-Mills theory. 

The second Casimir of the Yangian is computed in 
section 5. 
In conventional integrable models, the abelian Hamiltonians are 
found via the monodromy matrix, a useful holonomy that
is constructed from the R-matrix in terms of the local dynamical
variables $J^A_i$. 
In classical integrable models, expansion of the monodromy matrix
around $u=\infty$ in inverse powers of the spectral parameter gives the
Yangian generators and some non-local abelian charges.
The monodromy matrix has an isolated singularity
at $u=0$, which may be separated from the regular part of the expression
by rewriting the matrix as a product of three local factors
\refs{\Bernardthree, \faddeevfour}.
The trace of the monodromy can be shown then to have just one local
factor, when periodic boundary conditions are applied.
In this way, the reexpansion of the trace of the monodromy around $u=0$
gives the commuting local conserved quantities for the periodic chain. 

This procedure is somewhat modified in quantum integrable models,
due to the ambiguity in ordering the quantum operators.
It is convenient to multiply the $R$-matrix by $u$.
Again there are two sets of abelian conserved charges,
one is local and one is non-local.
Expanding the monodromy matrix in powers of the
spectral parameter gives the non-local Yangian generators and
non-local abelian charges \refs{\Bernardone}, where the
latter can also be derived by simply expanding the trace
of the monodromy matrix, the transfer matrix.
Expanding the logarithim of the transfer matrix
gives the local commuting Hamiltonians \refs{\Luschertwo - \grabowskitwo}.

But in the planar Yang-Mills theory, it is difficult to perform these
expansions and recover explicit expressions for the higher local
abelian conserved charges. We are thus led to present an alternative
derivation by finding them as Casimirs of the Yangian.

In section 6, the second Casimir is 
shown to annihilate the chiral primary states.
This is consistent with our experience that all one-loop corrections
are related to the one-loop anomalous dimension, which vanishes  
for chiral primary states. 

In section 7, we discuss how extending the Yangian to higher orders in
$g^2N$ addresses higher-loop integrability.
We have conjectured
in \refs{\DNW} that the Yangian generators can be defined
to all orders in $g^2 N$ via
$\tilde J^A = J^A + g^2 N J^A_2 +\ldots$, $\tilde Q^A = Q^A +  g^2 N
Q^A_2 + \ldots$, etc.$\,$, although $\tilde Q^A$ will not have the
simple bilinear form that it has at tree level.
The planar ${\cal N}=4$ Yang-Mills theory (SYM)
is radially quantized on $R\times S^3$, with the
Hamiltonian given by the dilatation operator $\tilde D$,
one of the ordinary symmetries of $PSU(2,2|4)$. The first spin chain
Hamiltonian $H_1$ is interpreted as the one-loop $g^2 N$
contribution to the dilatation operator, $D_2$.
The spin chain higher Hamiltonians $H_\kappa$, $\kappa >1$, which we
introduce in section 6 as Casimirs, 
are interpreted as one-loop contributions to other operators in SYM, 
which have zero tree-level contribution. They have been used to 
provide an understanding of
certain degeneracies of the one-loop anomalous dimension 
for charge-conjugated states \refs{\Beisertfour}. 

Their higher-loop contributions will be Casimirs of the higher-loop
Yangian charges,
but we do not calculate any higher-loop corrections here.
Since the R-matrix and monodromy satisfy algebraic constraints and
are derived from a universal R-matrix which is an element of the Yangian
algebra, the algebra is important for studying
any choice of boundary conditions. 

Appendix $A$ adapts some arguments developed for the $PSU(2,2|4)$
gauge theory to the spin $s$ $SU(2)$ chain.

\newsec{ $XXX_{1/2}$ Model}

We recall the role of Yangian symmetry in integrable models
by first considering the familiar quantum $XXX_{1/2}$ model,
the ordinary Heisenberg spin chain, to review and fix the nomenclature.  
For $L$ sites, its operators are
defined in a Hilbert space ${\cal H}_L =
\bigotimes_{i=1}^L h_i$. At each site $i$, the
local spin variable  $J^A_i$ is in the spin $1\over 2$ representation
of $SU(2)$ and acts as ${\sigma^A\over 2i}$
on a space  $h_i$ that is spanned by two states, corresponding 
to up or down spin. $\sigma^A$ are the Pauli matrices.
We have \eqn\lalg{[J^A_i, J^B_j] = \epsilon_{ABC} J^C_j \delta_{ij}\,.}
The spin chain Hamiltonian is
\eqn\ham{H = \beta\sum_{i=1}^L ( J^A_i J^A_{i+1} + {1\over 4})\,,}
where positive (negative) $\beta$ corresponds to the ferromagnetic 
(antiferromagnetic) case. For the moment, we do not impose
periodic boundary conditions. 
This makes it possible to construct the Yangian generators ${\cal J}^A_n$,
with $n= 0,\ldots, L$. The first two generators will be denoted as 
${\cal J}^A_0 = J^A, {\cal J}^A_1 = Q^A$, and ${\cal J}^A_n$
is an operator that acts on $n+1$ sites at a time
and arises from commutators of the $Q^A$.
The total spin variables
\eqn\spin{J^A = \sum_{i=1}^{L+1} J^A_i}
are the ordinary $SU(2)$ symmetry generators, with 
\eqn\comrel{[J^A,J^B] =  \epsilon_{ABC} J^C\,.}
The charges $J^A$ commute with the Hamiltonian
\eqn\symcom{[H, J^A]=0\,,} due to the $SU(2)$ symmetry of the model.
The bilocal Yangian generators are represented by
\eqn\nonloc{
Q^A = \epsilon_{ABC} \sum_{1\le i<j\le L+1} J^B_i J^C_j\,.}
They act on two sites at a time like the Hamiltonian,
but involve pairs that are not necessarily nearest neighbors,
as well as the group structure constants.
Their commutation with the ordinary symmetry generators is
\eqn\yangs{[J^A, Q^B] = \epsilon_{ABC} Q^C\,.}
We recall that their commutation with the Hamiltonian
is given by \eqn\edge{[H, Q^A] =  {1\over 2} \beta \,(J^A_1 - J^A_{L+1})\,,}
which can be shown by first considering a system of two spins, and proving 
\eqn\ledge{[H_{12}, Q_{12}^A] =  {1\over 2}\beta \,  q^A_{12} \,,}
where the two-body operators are 
$H_{i\, i+1} = \beta ( J^A_i J^A_{i+1} + {1\over 4} ) \,,$
$Q_{ij}^A = \epsilon_{ABC} J_i^B J_j^C\,,$
and $q^A_{ij}$ is the difference operator
\eqn\difo{q^A_{ij} = J^A_i - J^A_j\,.}
We remark that \ledge\ 
follows simply from the properties of the spin ${1\over 2}$
representation at each site,
since for $J^A_j = {\sigma^A_j\over 2i}$ we have 
\eqn\spm{\eqalign{[H_{12}, Q^A_{12}] &= \beta
\,\epsilon_{ABC} [ J^D_1 J^D_2, J^B_1 J^C_2]\cr
&= {1\over 2} \beta \,\, (J^A_1 - J^A_2)\,.\cr}}
For a chain of $L+1$ spins, 
the commutator $[H,Q^A]$ is given by 
\eqn\prffour{[H,Q^A]=\sum_{i=1}^L [H_{i,i+1},Q^A_{i,i+1}]
= {\beta\over 2}  \sum_{i=1}^L q^A_{i,i+1}= {1\over 2} \beta\, q^A,} where
$q^A=J^A_1-J^A_{L+1}$, since the cross terms 
{\it cf.}\refs{\DNW} vanish, 
\eqn\untu{0=\left[H_{i,i+1}\,,\,\sum_{j<k,(j,k)\not= (i,i+1)}
Q^A_{jk}\right].} This verifies the claim \edge\ ,
that for finite chains, the Yangian commutes with
the Hamiltonian $H$ up to edge effects. For chains of infinite length 
where we ignore the lattice total derivative, the Yangian is an
exact symmetry. 

For a finite chain with periodic boundary conditions
\eqn\bc{J_i^A = J_{i+L}^A\,,}
we also have $[H, Q^A] = 0$. We will see in section 6 that this implies 
the Casimir operators of the Yangian can still be defined for the periodic 
chain, even though the Yangian representation \nonloc\ cannot.  
These Casimirs are local and their commutator with $H$ vanishes.  
We note that \edge\ is the 
SU(2) spin ${1\over 2}$ analogue of the $[H, Q^A]$ commutator
which was derived for the $PSU(2,2|4)$ chain in \refs{\DNW}.
There an argument involving the properties of the two-particle modules
for the planar gauge theory was needed, whereas here
Pauli matrix identities suffice. 
In fact, as shown in \refs{\Minahan},
the Hamiltonian \ham\ can be used to
find the one-loop anomalous dimensions for a subsector of the 
gauge theory, with the ferromagnetic choice $\beta = {g^2N\over 4\pi^2}$ 
and the boundary conditions \bc\ .

$H$ is an integrable Hamiltonian. For periodic boundary conditions,
this means that it belongs to a set of $L$ commuting operators. 
In the thermodynamic limit ($L\rightarrow\infty$),
the abelian symmetry of the family of integrable Hamiltonians 
and the non-abelian Yangian symmetry become infinite-dimensional.
The abelian Hamiltonians are given by the transfer matrix, the
trace of the monodromy matrix, which also depends on $L$ sites, and
in turn is constructed from the R-matrix which depends on two sites.
The R-matrix satisfies a Yang-Baxter equation. 
This implies certain relations between the elements of the monodromy matrix
which can lead ultimately to the spectrum and correlation functions
of the model, via the algebraic Bethe ansatz \refs{\faddeev}.
We review some of its features with a view toward
introducing our Yangian methods. 

The R-matrix is defined on two sites and 
is a function of the spectral parameter $u$ 
\eqn\lax{R_{jm}(u) = (u + {i\over 2})
I_j\otimes I_m - J_j^A\otimes \sigma_m^A\,,}
where $J^A_i$ is the local spin variable introduced above, and
$\sigma^A_m$ are Pauli matrices defined at site $m$.
Our convention is $R_{jm}(0) = i P_{jm}$, where $P_{jm}$ permutes the
sites at $j$ and $m$:
\eqn\perm{P_{j,m}\equiv \half (I_j\otimes I_m
+ \sigma^A_j\otimes\sigma^A_m)\,.}
The non-abelian properties of the $R$-matrix \lax\
are given by the Yang-Baxter equation,
\eqn\ybe{R_{mn}(u-v) R_{im}(u) R_{in}(v) =
R_{in}(v)  R_{im}(u) R_{mn}(u-v)}
which is satisfied due to \lalg\ and the familiar properties of 
the Pauli matrices, 
and encodes the Yangian symmetry.
The sites $i,j$ are 
any points on the quantum chain; and $m,n$ label an auxiliary space.
The  R-matrix on two auxiliary sites has a similar form
with $J^A_m \equiv {\sigma_m^A\over 2i}$.
\lax\ is Yang's solution to the Yang-Baxter equation, and the
one relevant for the $XXX_{1\over 2}$ model. 
The monodromy matrix $T_m(u)$ defines the transport from site $1$ to site $L$,
and for convenience is defined in terms of a Lax operator
\eqn\rmatrix{L_{jm}(u) = R_{jm}(u - {i\over 2})\,,}
as
\eqn\mono{T_m(u) \equiv  L_{L,m}(u)\ldots L_{1,m}(u)\,.}
The shift in $u$ makes the Bethe Ansatz equations more symmetric, but
is not necessary to define monodromy.
For periodic boundary conditions, 
its trace, the transfer matrix, produces
commuting operators $M_\kappa$
\eqn\transfer{F(u) = \tr\, T_m(u) = 2 u^L + \sum_{\kappa=0}^{L-2}
M_\kappa u^\kappa \,,} where the trace is on the matrix 
at site $m$. The commuting operators 
defined by $M_\kappa$ are in general non-local. To access the {\it local}
abelian Hamiltonians, one must take derivatives of the logarithm of the
transfer matrix, 
\eqn\lt{H_\kappa  
\sim {{d^\kappa}\over du^\kappa} \ln F(u) |_{u={i\over 2}}}
in the fashion of keeping only connected graphs for Green functions
in a field theory\refs{\Luschertwo}. The $\sim$ defines $H_\kappa$ up
to overall multiplicative and additive constants. 
In particular, the first two local charges can be defined as 
$H_1$ given by \ham, corresponding to 
$H_1 = \beta ( -{i\over 2} {d\over du} \ln F(u) + {1\over 2} L) )|_{u=0}$,
and
\eqn\secondlocal{H_2 = \sum_{i=1}^L [P_{i,i+1}, P_{i+1,i+2}]\,.}

The monodromy also satisfies a Yang-Baxter relation
\eqn\monoybe{R_{mn}(u-v) T_{m}(u) T_{n}(v) =
T_{n}(v)  T_{m}(u) R_{mn}(u-v)\,,}
which implies that 
$F(u) F(v) = F(v) F(u)$, so that $[H_\kappa, H_{\kappa'}] = 0$.

The monodromy matrix is a polynomial in $u$, and
its components are the Yangian generators together with some non-local abelian
charges, all of which can still be defined 
since $T_m(u)$ involves only $L$ sites.
For {\it eg.}, for $L=2$ 
\eqn\monoyang{\eqalign{T_{m}(u) &= L_{2m}(u) L_{1m}(u) \cr
&= u^2 I_m
- u (J^A_1 + J^A_2) \sigma^A_m 
+ J_2^B J_1^C (\sigma^B\sigma^C)_m\cr
&= u^2 I_m 
- u J^A\sigma^A_m - i Q^A \sigma^A_m
+({1\over 2}J^A J^A +{3\over 4}) I_m\,.\cr}}
With a view toward extending the monodromy 
to higher-loops in $g^2N$, which we discuss in section 7,
one might inquire how direct is the relation between 
the Yang-Baxter equation \monoybe\ and  
the Yangian. 
We can show for $T_m(u)$ given by \monoyang\ , that
\monoybe\ follows immediately from the Yangian commutation relations
\comrel\ and \yangs\ and a two-site subsidiary condition
\eqn\extra{[Q^A, Q^B] = \epsilon_{ABC} ({1\over 2} J^D J^D J^C + {3\over 4}
J^C)\,.}
This extra condition \extra\ holds by inspection when
$J^A$ and $Q^A$ are in the representation given by \spin\ and \nonloc\ 
with $J^A_j = {\sigma^A_j\over 2i}$.
Given that the fundamental relation \monoybe\ holds for $L=2$, it follows 
for arbitrary $L$ due to the commutativity of  
R-matrices with no common sites, see {\it eg.}\refs{\faddeev}. 

Note that for $L>2$, the monodromy matrix will contain higher Yangian
charges ${\cal J}_n^A$, for $n= 0,\dots L-1$,
where ${\cal J}_0^A = J^A, {\cal J}_1^A = Q^A,\ldots$. For general $L$, 
the monodromy can be expressed in terms of the Yangian generators by
\eqn\mono{T_{ab}(u) = u^L \delta_{ab} + \sum_{n=0}^L u^{L-n}
t^{(n)}_{ab}\,,}
where $a,b$ label the indices of the matrix on the auxiliary site $m$ and
\eqn\crit{\eqalign{t^{(0)}_{ab} &= J^A T^A_{ab}\cr
t^{(1)}_{ab} &= Q^A T^A_{ab} + \half t^{(0)}_{ad}t^{(0)}_{db}
+ \alpha \delta_{ab} + \beta t^{(0)}_{ab}\,,}}
for some coefficients $\alpha, \beta$. The coefficients $t^{(n)}_{ab}$
for $n>1$ will involve the higher Yangian generators ${\cal J}^A_n$.
It can be shown for all $N\ge 2$,
when $J^A_i$ (and $T^A_{ab}$) 
are in the $N$-dimensional representation of $SU(N)$, that
the components of the monodromy satisfy commutation relations for
$[t^{(n)}_{ab}, t^{(m)}_{cd}]$ which are equivalent to the Yangian
defining relations \refs{\Bernardtwo}. The fact that 
the transfer matrix \mono\ satisfies the Yang-Baxter relation
\monoybe\ for some suitable $SU(N)$  $R$-matrix
again depends on the fact that  $J^A_i$ is in a 
special representation. 
So the extension of \mono\ and \crit\ to higher-loops in the gauge theory,
where the representation will change, is
not straightfoward, as we observe in section 7.
Nonetheless at the one-loop level, knowledge of
the monodromy matrix, whose logarithmic expansion gives 
the one-loop local Hamiltonians, is equivalent to 
the tree-level Yangian. 
 
\vskip5pt
 
\newsec{$XXX_{s}$ Model}
\vskip5pt

The purpose of this section is to derive the R-matrix and the Hamiltonian
for the $XXX_s$ model with methods similar to those used in \refs{\DNW}.
This will streamline the computations and also allow us to generalize
the results to $PSU(2,2|4)$. The $XXX_{s}$ model is
a spin chain where the local spin variables $J_i^A$ take values in
the $2s+1$-dimensional space ${\cal C}^{2s+1}$, and $s=0,{1\over 2},
1,\ldots$ is the spin labeling representations of $SU(2)$.
Following the literature \refs{\KRS,\faddeev}, 
we look for an R-matrix that 
leads to monodromy with integrable local Hamiltonians.
The equation for the R-matrix follows from the
the Yang-Baxter relation for a universal R-matrix ${\cal R}$
and acts in  ${\cal A}\otimes {\cal A}\otimes {\cal A}$,
when ${\cal A}$ is the Yangian algebra of $SU(2)$:
\eqn\aybr{{\cal R}_{12}{\cal R}_{32} {\cal R}_{31}
= {\cal R}_{31}{\cal R}_{32} {\cal R}_{12}\,.}
Since ${\cal R}$ is an element in ${\cal A}\otimes {\cal A}$,
its representation has the form 
\eqn\repr{R_{s' s}(u-v) = (\rho(s,u)\otimes\rho(s',v) )\,{\cal R}_{12}\,,}
where in general $s,s'$ label the spin of the $SU(2)$ representations 
at two sites \refs{\faddeev}. For spin $s$ at site $i$ and spin $1\over 2$
at site $m$, the representation acts as
\eqn\reprhalf{(I\otimes\rho(s,u)\otimes\rho(\half,v)) \,{\cal R}_{32}
= R_{im}(v-u)  = (v-u+{i\over 2})
I_i\otimes I_m - J_i^A\otimes \sigma_m^A\,.}
Using standard procedures \refs{\faddeev}, one applies the representation
$\rho(s,\lambda)\otimes \rho(s,\mu)\otimes \rho({1\over 2},\sigma)$
to \aybr\ to find a linear equation for $R_{ij}(\lambda)$:
\eqn\harp{\eqalign{
&R_{ij}(\lambda-\mu)\, ((\sigma-\mu + {i\over2}) 
I_i\otimes I_m - J_i^A\otimes\sigma_m^A )\,
((\sigma-\lambda +  {i\over2}) I_j\otimes I_m - J_j^B\otimes\sigma_m^B )\cr
&= ((\sigma-\lambda + {i\over2}) I_j\otimes I_m - J_j^B\otimes\sigma_m^B)\,
((\sigma-\mu + {i\over2}) I_i\otimes I_m - J_i^A\otimes\sigma_m^A)\, 
R_{ij}(\lambda-\mu)\,.}}
Here the spin variables on the quantum space $J_i^A, J_j^B$ 
are taking values in ${\cal C}^{2s+1}$.
A general representation $\rho(s_1,\lambda)\otimes \rho(s_2,\mu)
\otimes \rho(s_3,\sigma)$ acting on \aybr\
would mean that the single-site spin variables at the first, second, and
third sites are in the representations
$s_1, s_2, s_3$ respectively.
(Note the permutation of indices between \ybe\ and \harp.)
Now we diverge from the standard derivation.
Labeling the sites $i,j$ as $1,2$, and letting $\mu=0$,
we rewrite \harp\ in terms of the two-site Yangian generators as
\eqn\sharp{\eqalign{&
R_{12}(\lambda) ( (iQ^A + \lambda J_1^A -(\sigma +{i\over 2}) J^A)
\sigma_m^A + J_1^A J_2^A I_m )\cr
&=  ( ( -iQ^A + \lambda J_1^A -(\sigma +{i\over 2}) J^A)
\sigma_m^A + J_1^A J_2^A I_m ) R_{12}(\lambda)\,,}}
where in this section we write 
$J^A = J^A_1 + J^A_2$, $Q^A = \epsilon_{ABC} J^A_1 J^B_2$.
We will look for a solution where the R-matrix depends on
$J^A_1, J^A_2$ only through the Casimir, {\it i.e.}
$R_{12}(\lambda, J^A_1 J^A_2)$.
For spin s, the Casimir of the $SU(2)$ symmetry acting on 
a two-particle state is
\eqn\cas{J^A J^A = - 2s(s+1) I_1\otimes I_2  + 2 J^A_1 J^A_2 
= - J_{12}(J_{12}+1) \equiv - J (J+1)\,.} 
Let $V_S$ be the $2s+1$-dimensional space of one-particle states 
(the states at one spin site).
The tensor product $V_S\otimes V_S$
decomposes into irreducible representations of $SU(2)$, 
\eqn\decomp{V_S\otimes V_S = \bigoplus_{j=0}^{2s} V_j\,.}
The operator $J$ in \cas\ has eigenvalue $j$ acting on a 
two-particle state $|\kappa(j)\rangle$ that is contained in $V_j$,
and $j = 0,1,\dots 2s$.
Since $[J^A, J^D_1 J^D_2] = 0$, and $[R_{12}, J^D_1 J^D_2] = 0$,
\sharp\ becomes
\eqn\carp{R_{12}(\lambda) ( iQ^A + \lambda J_1^A) =
( -iQ^A + \lambda J_1^A) R_{12}(\lambda) \,,} or equivalently, since
$[R_{12}, J^A ] = 0$,
\eqn\carps{R_{12}(\lambda) ( 2iQ^A + \lambda (J_1^A - J_2^A) ) =
( -2iQ^A + \lambda (J_1^A - J_2^A)) R_{12}(\lambda) \,.}
To solve \carp\ , define a permutation operator $P_{ij}$
in ${\cal C}^{2s+1}\otimes {\cal C}^{2s+1}$ such that
\eqn\perma{P_{12} J^A_1 P_{12} = J^A_2\,,\qquad 
P_{12} J^A_2 P_{12} = J^A_1\,,\qquad\, P_{12} P_{12} = 1\,.}
(For $s={1\over 2}$, the permutation 
$P_{12}$ is given by \perm\ ).
We let $R_{12} (\lambda) = r(\lambda, J) P_{12}$,
and find that \carp\ reduces to 
\eqn\fcarp{ r(\lambda, J) ( 2iQ^A + \lambda q^A ) =
(2iQ^A - \lambda q^A ) r(\lambda, J) \,,}
with 
\eqn\dif{q^A = J^A_1 - J^A_2\,,}
since $P_{12} Q^A P_{12} = - Q^A$ and 
$P_{12} q^A P_{12} = - q^A$. 
Using the identity 
\eqn\iden{[J^D J^D, q^A ] = 4 Q^A\,,}
we can evaluate
\fcarp\ acting on a two-particle state $|\kappa(j)\rangle$ that is
contained in $V_j$ (and so has eigenvalues of $J^D J^D$ given above).
Then
\eqn\rsol{r(\lambda, J) 
({i\over 2} J^D J^D q^A - {i\over 2} q^A J^D J^D + \lambda q^A ) 
|\kappa(j)\rangle
= ({i\over 2} J^D J^D q^A - {i\over 2} q^A J^D J^D - \lambda q^A )
r(\lambda, J) |\kappa(j)\rangle\,.}
From \refs{\DNW}, we know that the action of $q^A$  
on a state in $V_j$ can be written as a linear combination 
of states in $V_{j-1}$ and $V_{j+1}$, {\it i.e.} for any
$|\kappa(j)\rangle \,\in V_j$ we have
\eqn\qsol{q^A  |\kappa(j)\rangle = |\chi^A(j-1)\rangle + |\rho^A(j+1)\rangle\,,}
where
$|\chi^A(j-1)\rangle \in V_{j-1}$ and 
$|\rho^A(j+1)\rangle \in V_{j+1}$. Actually in \refs{\DNW} we proved \qsol\ 
for the two-particle modules of the $PSU(2,2|4)$ gauge theory.
But on inspection \qsol\ also holds for an arbitrary spin $s$ representation
of $SU(2)$ due to the $SU(2)$ tensor product decomposition. See Appendix A. 
Using \cas, we have $J^D J^D |\kappa (j)\rangle = - j(j+1) |\kappa(j)\rangle$,
and computing \rsol\ we derive 
\eqn\rsola{\eqalign{&
r(\lambda, J)
({i\over 2} J^D J^D q^A - {i\over 2} q^A J^D J^D + \lambda q^A )
|\kappa(j)\rangle\cr
&=  r(\lambda, j-1) ( \lambda + i j )  |\chi^A(j-1)\rangle + r(\lambda, j+1) 
( \lambda - i (j+1) )  |\rho^A(j+1)\rangle \,,}}
and
\eqn\rsolb{\eqalign{&
({i\over 2} J^D J^D q^A - {i\over 2} q^A J^D J^D - \lambda q^A )
r(\lambda, J) |\kappa(j)\rangle\cr
&= (- \lambda + i j ) r(\lambda, j)  |\chi^A(j-1)\rangle
+ (- \lambda -i (j+1) ) r(\lambda, j)  |\rho^A(j+1)\rangle\,.}}
Equating the coefficients of $ |\chi^A(j-1)\rangle$ , 
we find
\eqn\fsol{ ( \lambda + i j ) r(\lambda, j-1) = (- \lambda + i j ) 
r(\lambda,j)\,,}
and an equivalent equation for the coefficients of  $ |\rho^A(j+1)\rangle$.
The solution to \fsol\ is
\eqn\euler{ r(\lambda, j) = {\Gamma(j + 1 - i\lambda)\over
\Gamma(j + 1 + i \lambda)}\,,}
so the R-matrix that satisfies \sharp\ is given by
\eqn\rmat{R_{12} (\lambda) = {\Gamma(J + 1 - i\lambda)\over
\Gamma(J + 1 + i \lambda)} \,P_{12}\,.}
This is the standard expression for the $R$-matrix whose monodromy
generates the local abelian charges for the $XXX_s$ model with 
periodic boundary conditions, but we have derived it here by considering how
the Yangian acts on it. 

The integrable Hamiltonian can now be found \refs{\faddeev, \KRS}
from the spectral invariants of the monodromy constructed from the
$R$-matrix in \rmat, 
\eqn\mon{T_f = R_{L\,f}(\lambda) R_{L-1\, f}(\lambda)\ldots
R_{1 f} (\lambda)\,,}
{\it i.e.} from the transfer matrix $F_f(\lambda) = \tr_f T_f(\lambda)$
which satisfies ${[F_f(\lambda), F_f(\mu)] = 0\,.}$ Hence 
\eqn\inth{H \sim - i {d\over d\lambda} \ln F_f(\lambda) |_{\lambda =0}= 
- \sum_{i=1}^L  2\psi(J_{i,i+1} + 1)\,,}
since 
$i  {d\over d\lambda} \ln r(J_{j,j+1},\lambda) |_{\lambda =0}
=  2 \psi(J_{j,j+1} +1)\,$, where $\psi(x) = {d\over dx}\ln \Gamma(x)$.

Alternatively, we will show that we can derive the Hamiltonian {\it directly}
from the Yangian, without going through the $R$-matrix,
by requiring that $H_{12}$ 
be some function of the Casimir $J$ that satisfies 
\eqn\check{[H_{12}(J), Q^A] = q^A\,.}
Evaluating \check\ via \iden, we have
\eqn\checka{\eqalign{&[H_{12}(J), Q^A] |\kappa(j)\rangle\cr 
&= {1\over 4} ( H_{12}(J) J^DJ^D q^A -  H_{12}(J) q^A J^D J^D
-  J^D J^D q^A  H_{12}(J)  + q^A J^D J^D  H_{12}(J))  |\kappa(j)\rangle\cr
&= - {1\over 2} ( j (H_{12}(j) - H_{12}(j-1) ) |\chi^A(j-1)\rangle + 
(j+1) (H_{12}(j+1) - H_{12}(j)) |\rho^A(j+1)\rangle)\cr
&= (|\chi^A(j-1)\rangle + |\rho^A(j+1)\rangle) =  q^A  |\kappa(j)\rangle\,,}}
which requires
\eqn\heq{j \,(H_{12}(j) - H_{12}(j-1)) = - 2\,.}
We can solve \heq\ by adding it together for different
values of integer $j$ to find 
\eqn\hsol{\half (H_{12}(j) - H_{12}(0)) = - \sum_{n=1}^j {1\over n}\,,}
whose solution is proportional to the digamma function up to an 
additive constant,
\eqn\dig{H_{12}(J) = - 2 (\psi(J+1) - \psi(1) )\,.}
The overall minus sign in \dig\ 
results from the normalization \cas, and
corresponds to the ferromagnetic case with $\beta = 2$.
Following \refs{\faddeev}, it is often useful to think of $\psi(J+1)$ as
a polynomial in $J^A_1 J^A_2$, given by $f(J^A_1 J^A_2)$, which is
equal to $\psi(J+1)$ at the eigenvalues of $J^A_1 J^A_2$.
Using the value of $\psi(k+1) \equiv \sum_{n=1}^k {1\over n} - \gamma$ 
for certain values of $k$ (the non-negative integers), one can write a
Lagrange interpolating polynomial for
$\psi(J +1)$ in terms of 
the nearest neighbor pair $J_1^A J_2^A$, which has eigenvalues 
$x_j = s(s+1) - {1\over 2} j (j+1)$,
\eqn\nn{\psi(J + 1) = \sum_{k=0}^{2s} c_k (J_1^A J_2^A )^k
= f_{2s} (J_1^A J_2^A)\,,}
and where the polynomial $ f_{2s} ( x)$ is
\eqn\polyn{ f_{2s} ( x) = \sum_{k=0}^{2s}\, 
(\, \psi(k+1) \, \prod_{j = 0, j\ne k}^{2s}
{{x-x_j}\over {x_k-x_j}}\,)\,.}
In particular, for $s=\half$, $f_{2s}(x) = - x+{3\over 4}-\gamma$ so
we regain $H_{12} = 2 J^A_1 J^A_2$, up to a constant.

When $s\ne 0,{1\over 2}, 1, \ldots$, 
the spin model can have local spin variables $J^A_i$ taking values in
an infinite-dimensional space: $J^A_i$ acts on $V_S$ where
$V_S$ is an infinite-dimensional spin $s$-representation
and $V_S \otimes V_S = \sum_{j=0}^{\infty} V_j$, and
$V_j$ are the two-site infinite-dimensional irreducible representations.
The Hamiltonian \dig\ and R-matrix \rmat\ are continued to general $s$,
for any real form of $SU(2)$.
As discussed in  \refs{\Beisertfour, \Beisertone},
a non-compact $SL(2)$ subsector 
of the four-dimensional planar superconformal Yang-Mills theory leads to a 
one-loop anomalous dimension operator $H_{12}$
whose eigenvalues can be identified with 
$\sum_{j=0}^\infty (2\psi (j+1) +2\gamma)$.
The action of the local spin variable $J_i^A$ on
the one-particle states corresponds to $s=-{1\over 2}$,
and one concludes that in this subsector
of states, the one-loop anomalous dimension operator is 
the $XXX_{-{1\over 2}}$ Hamiltonian for the Lie algebra $SU(1,1)$.
In this case, \cas\ becomes $J^A J^A = {1\over 2} I_1\otimes I_2
+ 2 J^A_1 J^A_2 = - J(J+1)$, and the spin of $V_j$ is $-1-j$.

\vskip5pt
\newsec{$PSU(2,2|4)$ Integrable Spin Chain}
\vskip5pt
Now we consider a spin model 
with local spin variables $J^A_i$ 
satisfying the $PSU(2,2|4)$ superalgebra with structure
constants $f_{ABC}$,
\eqn\newalg{[J^A_i, J^B_j\} = f^{AB}_C J^C_j \delta_{ij}\,,}
in the representation where $J^A_i$ acts on the space $V_F$.
$V_F$ is spanned by 
one-particle states in free $D=4$, $N=4$ superconformal Yang-Mills theory
radially quantized on ${\bf R}\times S^3$, and is an 
infinite-dimensional representation of 
$PSU(2,2|4)$ with $J^A_1 J^A_1 \, V_F = 0$.

The two-particle Casimir, with $J^A = J^A_1 + J^A_2$, is given by
\eqn\casa{J^A J^A = J(J+1)\,.}
The tensor product decomposition is
\eqn\crow{V_F\otimes V_F = \bigoplus _{j=0}^\infty V_j\,,} so
the eigenvalue of $J$ takes values $j=0,1,2,\dots,\infty$.
$V_j$ are $PSU(2,2|4)$ irreducible infinite-dimensional 
representations corresponding to the two-particle states.
Note that \casa\  holds in the two-particle sector of $PSU(2,2|4)$
$SYM$ theory, as though the algebra were $SU(2)$.
We have chosen a positive normalization in \casa,
which is consistent with \cas, since the  
the $PSU(2,2|4)$ Casimir contains the negative of the generators
$R^a_b R^b_a$ of the $SU(2)$ Casimir
\refs{\Beisertone, \DNW}.

Following the discussion in the previous section, we can
derive the form of the one-loop  $PSU(2,2|4)$ Hamiltonian from
the tree level Yangian generator
\eqn\yangps{Q^A = f^A_{BC}\sum_{i<j} J_i^B J_i^C\,,}
by again requiring that the density $H_{12}(J)$ satisfies
\eqn\acheck{[H_{12}(J), Q_{12}^A] = q_{12}^A\,,}
for $q_{12}^A = J_1^A - J^A_2$.
Using \iden, \qsol\ and \casa, we find 
\eqn\bcheck{\eqalign{&[H_{12}(J), Q^A] |\kappa(j)\rangle\cr
&= {1\over 4} ( H_{12}(J) J^DJ^D q^A -  H_{12}(J) q^A J^D J^D
-  J^D J^D q^A  H_{12}(J)  + q^A J^D J^D  H_{12}(J) )|\kappa(j)\rangle\cr
&= {1\over 2} ( j (H_{12}(j) - H_{12}(j-1) ) |\chi^A(j-1)\rangle + 
(j+1) (H_{12}(j+1) - H_{12}(j)) |\rho^A(j+1)\rangle)\cr
&= (|\chi^A(j-1)\rangle + |\rho^A(j+1)\rangle) =  q^A  |\kappa(j)\rangle\,,}}
where now the result \qsol\ applies to the two-particle modules
of the $PSU(2,2|4)$ gauge theory, as we already proved in \refs{\DNW}.
Solving
\bcheck\ requires
\eqn\aheq{j \,(H_{12}(j) - H_{12}(j-1)) =  2\,,}
so adding \aheq\  for different values of integer $j$, we find
$\half (H_{12}(j) - H_{12}(0)) =  \sum_{n=1}^j {1\over n}$,
and \eqn\adig{H_{12}(J) = 2 (\psi(J+1) - \psi(1) )\,.}
This is the same Hamiltonian density derived from one-loop Feynman
graphs in the superconformal gauge theory \refs{\Beisertone,
\Beisertfour}.
For a chain of more than two spins, $H$ is a sum of nearest neighbor
terms $H_{i,i+1}$. 

Now we extend the derivation of the R-matrix in the previous section 
to $PSU(2,2|4)$. Therefore we will introduce two appropriate representations,
one for the auxiliary space and one for the quantum space.
For our discussion, we will not need to distinguish between
$PSU(4|4)$ and its real form, since we are using algebraic procedures.
It will be useful to first consider the extended group $U(4|4)$.
As in \refs{\cinn}, we can define a single site representation given by 
the $4|4$ representation of the group $U(4|4)$,
which has two extra $U(1)$ generators $K$ and $R$. We will call this
representation $T_A$. The $U(1)$ R-symmetry
acts on the representation but is not a symmetry of the gauge theory,
and the $T_K$ generator acts by $T_K=0$. 
We will also need the metric for $U(4|4)$ which we take to be
$g_{AB}=\half \Str \,T_AT_B$, where we now let $1\le A,B\le 64$. 
We use the conventions for the supertraces in \refs{\cinn}.
So $g_{KK}=g_{RR}=0$,
$g_{KR}\not=0$;
and when $A\ne K,R$ then $g_{AB} = 2 \delta_{AB}$ and $g_{KA} = g_{RA} = 0$.
This metric
will be used to raise and lower the ``$A$'' index of the Lie algebra
generators. The $4|4$ representation $T_A$ has
$[T_A,T_B] = f_{AB}^C T_C$, where $f_{AB}^C$ are now the $U(4|4)$
structure constants. 
Then $f^A_{KB}=0$ for all $A,B$ since $K$ is central and commutes
with everything, and 
\eqn\fire{f^R_{AB}=0} for all $A,B$, since the $U(1)$
$R$-symmetry generator $R$ never appears on the right hand side of
the commutation relations. The metric is used to find
$f_A^{RB}=0=f_K^{AB}$.

The second representation is defined by interpreting the infinite-dimensional
module $V_F$ as a representation of the extended group $U(4|4)$ as follows.
We include the chirality operator $B$ and the central charge $C$
in the set $J^A_i$ to form generators of $U(4|4)$.
On the states in $V_F$, $C$ acts as $0$, and $B$ gives the chiral charge. 


We consider an abstract Yang-Baxter relation for a 
universal R-matrix ${\cal R}$,
which is defined as an element in ${\cal A}\otimes {\cal A}$,
when ${\cal A}$ is the Yangian algebra of $U(4|4)$. It is given by
\eqn\aybra{{\cal R}_{12}{\cal R}_{32} {\cal R}_{31}
= {\cal R}_{31}{\cal R}_{32} {\cal R}_{12}\,.}
We choose $s'$ to denote $T_{A\,m}$ given in terms of the
$4|4$ representation of $U(4|4)$,
and $s$ to denote that $J^A_i$ acts on $V_F$. 
Then the analogue of \reprhalf\ is 
\eqn\news{(I\otimes\rho(s,\lambda)\otimes \rho(s',\sigma)) {\cal R}_{32}
= R_{jm} (\sigma-\lambda) = (\sigma-\lambda + {i\over 2})
I_j\otimes I_m  + \eta J_j^A\otimes T_{A\,m}\,,}
where $\eta$ is an arbitrary normalization constant.
Applying the representation
$\rho(s,\lambda)\otimes \rho(s,\mu)\otimes \rho(s',\sigma)$
to \aybra\ we write a linear equation for $R_{ij}(\lambda)$:
\eqn\harp{\eqalign{&R_{ij}(\lambda-\mu) ((\sigma-\mu) I_i\otimes I_m 
+ \eta J_i^A\otimes T_{A\,m})
((\sigma-\lambda) I_i\otimes I_m + \eta J_j^B \otimes T_{B\,m} )\cr
&=  ((\sigma-\lambda) I_i\otimes I_m + \eta  J_j^B\otimes T_{B\, m} )
((\sigma-\mu) I_i\otimes I_m + \eta J_i^A\otimes T_{A\,m}) 
R_{ij}(\lambda-\mu)\,.\cr}}
As in the previous section, we set $\mu = 0$, and write $i,j$ as $1,2$.
Then 
\eqn\hobo{\eqalign{&R_{12}(\lambda) ( (\sigma J^A - \lambda J^A_1) T_{Am} 
+\eta J^A_1 J^B_2 (T_A T_B)_m)\cr
&= (  ( \sigma J^A - \lambda J^A_1) T_{Am} +\eta J^A_1 J^B_2 (T_B T_A)_m)
\, R_{12}(\lambda)\,.}}
For a single site $i$, let $J^a_b \equiv J^A_i (T_A)^a_b$, where
$a,b$ label the matrix indices at the site $m$.
We will use the property that the representation $V_F$ satisfies the
criterion at a given site $i$, 
\eqn\crit{J^a_b  J^b_c = \alpha J^a_c}
modulo the identity $\delta_{ac}$, for some proportionality constant
$\alpha$ \refs{\DNW,\cinn}. 
This occurs for any representation $M$, when $M\otimes \bar M$ contains
the adjoint only once, such as the $n$-dimensional representation of $SU(n)$.
Note that this criterion \crit\ was needed in order for
\yangps\ to satisy the Yangian Serre relations \refs{\cinn}.
Normalizing $[T_A, T_B\} =  f_{AB}^C T_C$, with $f_{AB}^C$ here given by the
$U(4|4)$ structure constants, and using \crit, we have 
\eqn\rtimesr{\eqalign{J^A_1 J^B_2 (T_A T_B)^a_b
&= \half J^A_1 J^B_2 [T_A, T_B\}^a_b
+ \half (J^A_1 J^B_2 + J^A_2 J^B_1) (T_AT_B)^a_b\cr
&=  \half Q^A T_A + \half J^A J^B T_A T_B - \half (J^A_1 J^B_1 + J^A_2 J^B_2)
T_A T_B\cr
&= \half Q^A T_A +  \half J^A J^B T_A T_B - \half\alpha J^A T_A\,,\cr}} 
where here $Q^A$ denotes the representation of the $U(4|4)$ Yangian on 
two sites \eqn\gall{\eqalign{
Q^C = f^C_{AB} J_1^A J_2^B\,.\cr}}
So \hobo\ reduces to 
\eqn\hoho{\eqalign{&R_{12}(\lambda) 
((\sigma J^A - \lambda J^A_1) T_{A\,m} +
\eta (\half Q^A T_{A\,m} + \half J^A J^B (T_A T_B)_m + \alpha J^A T_{A\,m}))\cr
&= (  ( \sigma J^A - \lambda J^A_1) T_{A\,m} +\eta ( -\half Q^A T_{A\,m}
+  \half J^A J^B (T_A T_B)_m + \alpha J^A T_{A\,m}))
\, R_{12}(\lambda)\,.\cr}}
Using the reasoning from the previous section, we assume
the $R$-matrix depends on $J^A_1, J^A_2$ only through the $U(4|4)$ two-particle
Casimir $J^A J^A$, so that $[R_{12}(\lambda), J^A ] = 0$ and 
\hoho\ reduces to 
\eqn\sharp{R_{12}(\lambda) ( \eta Q^A - \lambda q^A ) T_A=
- ( \eta Q^A + \lambda q^A) T_A R_{12}(\lambda) \,,}
with \eqn\dif{q^A = J^A_1 - J^A_2\,.}
We look for a solution that acts on each irreducible $V_j$ module in \crow,
\eqn\ssolr{R_{12} (\lambda) = r(\lambda, J) P_{12}}
where $P_{12}$ permutes the fields at sites $12$ in the module $V_j$.
$J$ is the $PSU(4|4)$ Casimir defined via \casa, since
the difference between the $U(4|4)$ Casimir and the $PSU(2,2|4)$ Casimir
is only $BC$, and the central charge $C$ acts as $0$ on $V_F$.
We argue that  \sharp\ becomes
\eqn\fcarp{ r(\lambda, J) (- \eta Q^A + \lambda q^A ) =
- (\eta Q^A + \lambda q^A ) r(\lambda, J) \,,}
for $A$ restricted to the $PSU(4|4)$ indices $1\le A\le 62$,
since  $K = T_K = 0$ in this representation, and 
$Q^R=0$ from \gall\ and \fire.  Also, $q^R = 0$, since $J^R_i$ is $C$, which 
acts as zero. 

Using \iden, \qsol\ for $PSU(4|4)$, and \casa, we make the now familiar
argument and let
\fcarp\ act on a two-particle state $|\kappa (j)\rangle$ contained in
the module $V_j$ defined in \crow,
\eqn\sunny{r(\lambda, J)
( - {\eta\over 4} J^D J^D q^A + {\eta\over 4} q^A J^D J^D + \lambda q^A )
|\kappa(j)\rangle
=  - ({\eta\over 4}  J^D J^D q^A - {\eta\over 4}  q^A J^D J^D + \lambda q^A )
r(\lambda, J) |\kappa(j)\rangle\,.}
to find
\eqn\rsolar{\eqalign{& 
r(\lambda, J)
(-{\eta\over 4}  J^D J^D q^A + {\eta\over 4}  q^A J^D J^D + \lambda q^A )
|\kappa(j)\rangle\cr 
&=  r(\lambda, j-1) ( \lambda + {\eta\over 2} j )  
|\chi^A(j-1)\rangle + r(\lambda, j+1)
( \lambda - {\eta\over 2} (j+1) )  |\rho^A(j+1)\rangle \,,}}
and
\eqn\rsolbr{\eqalign{&
(- {\eta\over 4} J^D J^D q^A + {\eta\over 4}  q^A J^D J^D - \lambda q^A )
r(\lambda, J) |\kappa(j)\rangle\cr
&= (- \lambda + {\eta\over 2} j ) r(\lambda, j)  |\chi^A(j-1)\rangle
+ (- \lambda -{\eta\over 2} (j+1) ) r(\lambda, j)  |\rho^A(j+1)\rangle\,.}}
Equating the coefficients of $ |\chi^A(j-1)\rangle$ , or equivalently those
of $|\rho^A(j+1)\rangle$,
we have 
\eqn\gsol{ ( {\eta\over 2} j + \lambda ) r(\lambda, j-1) 
= ({\eta\over 2} j - \lambda ) r(\lambda,j)\,,}
so the R-matrix 
that satisfies \harp\ is
\eqn\srmat{R_{12} (\lambda) = {\Gamma(J + 1 + {2\over \eta} \lambda)\over
\Gamma(J + 1 - {2\over\eta} \lambda)} \,P_{12}\,.} 
Since the values of the permutation are $P_{12} = (-1)^j$ on each $V_j$
\refs{\DNW}, we find that the R-matrix acts on 
$V_F\otimes V_F$ as ${\bf R}_{12}(\lambda)$ given by 
\eqn\banana{\eqalign{&{\bf R}_{12}(\lambda) 
= (-1)^J  {\Gamma(J + 1 + {2\over \eta} \lambda)\over
\Gamma(J + 1 - {2\over\eta} \lambda)}\cr
&= \sum_{j=0} (-1)^j  {\Gamma(j + 1 + {2\over \eta} \lambda)\over
\Gamma(j + 1 - {2\over\eta} \lambda)} P_{12j}\,,\cr}}
where $P_{12j}$ projects the fields at positions $i,i+1$ to
the module $V_j$.
Forming the transfer matrix, 
\eqn\TM{F_f(\lambda) = \tr_f  R_{L\,f}(\lambda) R_{L-1\, f}(\lambda)\ldots
R_{1 f} (\lambda)\,}
we can find the Hamiltonian again,
\eqn\hamagain{H =  \sum_{i=1}^L  2(\psi(J_{i,i+1} + 1)
-\psi(1)) \sim  {d\over d\lambda} \ln F_f(\lambda) |_{\lambda =0}\,.}

Up to multiplicative factors independent of $J$, 
the R-matrix \banana\ is the same as that found by extrapolating from a 
subsector of the gauge theory and assuming uniqueness \refs{\Beiserttwo}. 
This is seen easily from the reflection formulas
$\Gamma(z) \Gamma(1-z) = {\pi\over \sin{\pi z}}$ and
$\psi(1-z) = \psi(z) + \pi \cot{\pi z}$, and 
the integer eigenvalues of $J$.

Since \aybra\ is an abstract equation that holds in
${\cal A}\otimes {\cal A}\otimes {\cal A}$,
when ${\cal A}$ is the Yangian algebra of $U(4|4)$,
we could permute the indices on \aybra\ to find
\eqn\gosh{{\cal R}_{12}{\cal R}_{13} {\cal R}_{23}
= {\cal R}_{23}{\cal R}_{13} {\cal R}_{12}\,,}
which for the representation 
$\rho(s',\lambda)\otimes \rho(s',\mu)\otimes \rho(s,\sigma)$
results in
\eqn\pepper{R_{mn}(\lambda - \mu ) R_{im}(\lambda) R_{in}(\mu) =
R_{in}(\mu )  R_{im}(\lambda) R_{mn}(\lambda - \mu)\,.}
This is similar to obtaining \ybe\ in the $XXX$ model 
\refs{\faddeev}. It would be interesting to solve
\pepper\ for $R_{mn}(\lambda)$ in $PSU(4|4)$, but we do not do that here. 

\newsec{Commuting Hamiltonians for the Periodic Chain as
Casimirs of the $PSU(2,2|4)$ Yangian}

In this section we show how the bilocal Yangian generator, although itself 
not defined 
for periodic boundary conditions, can be used to find the
Casimirs for the periodic chain. 
We recall that the Hamiltonian \refs{\Beisertone} 
\eqn\fir{H = \sum_{i=1}^L  2 \,(\psi (J_{i, i+1} + 1) - \psi(1))\,,}
is the first Casimir operator of the $PSU(2,2|4)$ Yangian \refs{\DNW},
where $J_{i, i+1}$ is the two-site
quadratic Casimir of the ordinary symmetry $PSU(2,2|4)$ given by
\eqn\casi{(J^A_i + J^A_{i+1}) (J^A_i + J^A_{i+1})
= J_{i, i+1}  (J_{i, i+1} + 1)\,.}

We will demonstrate below 
that the second Yangian Casimir is
\eqn\casb{U = \sum_{i=1}^L ( \psi (J_{i, i+1}+ 1) \psi (J_{i+1, i+2} + 1)
- \psi (J_{i+1, i+2} + 1)  \psi (J_{i, i+1} + 1) )\,.}
We note that $U$ of \casb\ reduces to $H_2$ in \secondlocal\ 
for the $SU(2)_{1\over2}$
subsector of the gauge theory.
$U$ acts on three adjacent sites at a time, and
can be defined for a chain of three or more independent sites.

We have periodic boundary conditions
$J^A_i = J^A_{i+L}$, with $L$  independent sites, and as usual
\eqn\ords{J^A = \sum_{i} J^A_i\,,} 
where the index $i$ runs over the number of independent sites. 
Since $H$ and $U$ are functions of the ordinary Casimir $J_{i,i+1}$,
then
\eqn\ccom{[H, J^A] = 0\,,\qquad [U, J^A] = 0\,.}
We will now show that $U$ is a Casimir of the Yangian. As a first step,
we review why $H$ is a Casimir of the Yangian.
We start with an open chain 
$\bar H = \sum_{i=1}^L  2(\psi (J_{i\, i+1} + 1)- \psi(1))\,,$
where we do not have periodic boundary conditions,
and there are $L+1$ independent sites. 
The Yangian is well defined, with
\eqn\yangen{Q^A = f^A_{BC} \sum_{i<j} J^B_i J^C_j  \,,}
and the $i,j$ indices run over the number of independent sites, as long
as $i<j$.
We compute the commutation relation of $\bar H$ with the Yangian as
\eqn\hyangen{\eqalign{[\bar H, J^A ] &= 0\,,\qquad
[\bar H, Q^A] = q^A_{1\,L+1} = ( J^A_1 - J^A_{L+1})\,,\cr}}
{\it i.e.} 
$\bar H$ commutes with the Yangian up to the edge effect $q^A_{1\, L+1}$.
To prove \hyangen\ in \refs{\DNW}, we used, for $1\le i\le L$, that
\eqn\olf{\eqalign{
[ H_{i, i+1}, Q^A] &\equiv [2 \psi (J_{i, i+1} + 1) , Q^A] \cr
&=  [2 \psi (J_{i, i+1} + 1) ,  f^A_{BC} 
J^B_i J^C_{i+1} ] = q^A_{i,i+1}\,,\cr}}
and 
\eqn\oaf{\eqalign{
[ H_{i, i+1}, J^A] &\equiv [\psi (J_{i, i+1} + 1) , J^A]\cr
&=[\psi (J_{i, i+1} + 1) , J_i^A + J^A_{i+1}] = 0\,,}}
here the second lines in \olf\ and \oaf\ are due to the absence of 
cross terms {\it c.f.}\untu.
As an aside, notice that inversely, 
given the Yangian charge \yangen, then \olf\
could be used to derive $\bar H$ as in \bcheck-\adig.

Now we compute similar quantities for the second Casimir $U$.
Let 
$\bar U =\break 
\sum_{i=1}^L  ( \psi (J_{i, i+1}+ 1) \psi (J_{i+1, i+2} + 1)
- \psi (J_{i+1, i+2} + 1)  \psi (J_{i, i+1} + 1) )$ be an open chain
version of \casb,
where $\bar U$ is defined on $L+2$ independent sites. 
Since $\bar U$ is just a product of the $\psi$'s,
we can show 
\eqn\olaf{\eqalign{[\bar U, J^A] &= 0\,,\cr
[\bar U, Q^A] &= [\bar U, f^A_{BC} \sum_{1\le i\le j\le L+2} 
J^B_i J^C_j ]\cr
&=- [ q^A_{23}, \psi (J_{12} + 1) ] 
+ [q^A_{L,L+1}, \psi (J_{L+1, L+2}\, + 1) ]\cr
&= - [J^A_2,  \psi (J_{12} + 1) ] -  [J^A_{L+1}, \psi(J_{L+1, L+2}\, + 1) ]\cr
&= [J^A_1, \psi (J_{12} + 1) ] -  [J^A_{L+1}, \psi(J_{L+1, L+2}\, + 1) ]\cr
&= [(J^A_1 - J^A_{L+1}),  \psi (J_{12} + 1)] + [(J^A_1 - J^A_{L+1}), 
\psi(J_{L+1, L+2}\, + 1) ]\cr 
&= [ q^A_{1\,L+1},  (\psi (J_{12} + 1) + \psi(J_{L+1, L+2}\, + 1)) ]\,,\cr}}
where here the commutator is non-trivial for $L+2$ sites of the Yangian.
Note that \hyangen\ also 
holds for $J^A$ and $Q^A$ defined on $L+1$ or more sites, {\it eg.}
\eqn\ludwig{[\bar H, Q^A] = [ \bar H,  f^A_{BC} \sum_{1\le i\le j\le L+1} 
J^B_i J^C_j ] = [\bar H,  f^A_{BC} \sum_{1\le i\le j\le L+2} 
J^B_i J^C_j ] = q^A_{1\,L+1}\,.}
This follows from 
\eqn\oldy{[\bar H,  f^A_{BC} \sum_{i=1}^{L+1} J^B_i J^C_{L+2} ] = 0}
since  $\sum_{i=1}^{L+1} J^B_i$ is the $PSU(2,2|4)$ generator of the
$L+1$ spin system, and so commutes with $\bar H$.
To derive \olaf, we employ identities such as
$[q_{12}^A, \psi(J_{23} + 1)] - [q_{34}^A, \psi(J_{23} + 1)] =
- [(J^A_2 + J^A_3),  \psi(J_{23} + 1)] = 0$, that follow from 
\newalg\ and \casi. 
From \olaf,
the commutator of $\bar U$ with the Yangian is zero up 
to edge effects. 

We now identify the periodic chain expressions $H$ and $U$ 
as Casimirs of the Yangian by considering the commutators 
\hyangen\ and  \olaf. These commutators involve 
the open chain versions $\bar H$ and $\bar U$, where the Yangian makes sense.
Clearly we cannot impose periodic boundary
conditions \bc\ before performing the commutators, since 
$Q^A$
would not be defined.
But, if {\it after} performing the commutators in \hyangen, \olaf, 
we let $J^A_1 = J^A_{L+1}$, then the difference operator $q^A_{1\,L+1}$ 
vanishes, and the commutators are all zero. Then we find the periodic chain
expressions $H$ and $U$ in \fir\ and \casb, by
simply imposing the periodic boundary conditions \bc\ on 
$\bar H$ and $\bar U$. 
As an example, for $L = 3$, one finds
$\bar H = 2 (\psi(J_{12}+1)-\psi(1) + \psi(J_{23}+1)-\psi(1) +
\psi(J_{34}+1)-\psi(1))$. Setting $J^A_1 = J^A_4$ in $\bar H$, we recover 
$H =  2 (\psi(J_{12}+1)-\psi(1) + \psi(J_{23}+1)-\psi(1) +
\psi(J_{31}+1)-\psi(1))$, which is \fir\ for $L = 3$.
Similarly, $\bar U = [\psi(J_{12}),  \psi(J_{23})]
+ [\psi(J_{23}),  \psi(J_{34})] +  [\psi(J_{34}),  \psi(J_{45})]$,
and imposing  $J^A_1 = J^A_4$,  $J^A_2 = J^A_5$, we regain
the periodic chain expression $U =  [\psi(J_{12}),  \psi(J_{23})]
+  [\psi(J_{23}),  \psi(J_{31})] +  [\psi(J_{31}),  \psi(J_{12})]$
which is \casb\ for  $L = 3$.
It is in this sense that $H$ and $U$ are Casimirs of the Yangian.
\footnote*{In this paper, we are referring
to a set of local operators as Casimirs of the Yangian.}

These arguments are evidence of a commuting family of operators
defined by \TM, 
where $[F_f(\lambda), J^A] = 0\,,$ and $[F_f(\lambda), F_f(\mu)] = 0\,.$
Since $H$ and $U$ are both elements in the expansion of $F_f(\lambda)$,
it follows as usual that 
\eqn\nclo{[H,U] = 0\,.} 
A direct check 
that $[H,U]=0$ using the expressions in \fir\ and \casb\ 
involves the representations of $J_{i,i+1}$ and would be much harder.

Note that our ``open chain'' expressions do not commute, that is 
${[\bar H, \bar U] \ne 0\,}$,
and therefore should not 
be confused with commuting Hamiltonians for the open chain,
which vanish when the number of adjacent sites is odd \refs{\eks, \gm}.

As $L\rightarrow\infty$ , there will be a infinite number of the 
local abelian Hamiltonians for the periodic chain $H, U, \dots$,  
which are related to higher nested commutators
of the $\psi(J_{i,i+1} + 1)$.  Each of these will be Casimirs of the
Yangian. Since the Yangian $Y(G)$ has a basis  
${\cal J}_n^A$ where ${\cal J}_0^A=J^A$, ${\cal J}_1^A=Q^A$,
and ${\cal J}_n^A$ arises in 
commutators of the $Q^A$'s, any Casimir of $J^A$ and $Q^A$ is also a
Casimir of $Y(G)$.


\newsec
{Action of the Casimirs on the Chiral Primary States}

Like the Hamiltonian $H$, the second Yangian Casimir $U$ also annihilates
the chiral primary states, 
as we will demonstrate in this section. 
The lowest components of the chiral 
primary representations are built 
only from the scalar fields $\phi^I$, where $1\le I\le 6$.
For $L$ independent sites, 
this is the symmetric traceless product of $L$ {\bf6}'s,
where traceless representations are defined by those which give zero when any 
two indices are contracted, {\it eg.} \refs{\Aharony}.
So for $L = 2,3$, we have 
\eqn\candy{|\lambda_2\rangle = 
\phi^I\phi^J + \phi^J\phi^I - {1\over 3}\delta^{IJ}\phi^M\phi^M\,,}
\eqn\oscar{\eqalign{|\lambda_3\rangle&
=\phi^I\phi^J\phi^K + \phi^I\phi^K\phi^J + \phi^J\phi^I\phi^K 
+ \phi^J\phi^K\phi^I + \phi^K\phi^I\phi^J + \phi^K\phi^J\phi^I\cr
&\hskip6pt-{1\over 4} \delta^{IJ} (\phi^M\phi^M\phi^K + \phi^M\phi^K\phi^M
+ \phi^K\phi^M\phi^M)\cr
&\hskip6pt-{1\over 4} \delta^{IK} (\phi^M\phi^M\phi^J + \phi^M\phi^J\phi^M
+ \phi^J\phi^M\phi^M)\cr
&\hskip6pt-{1\over 4} \delta^{JK} (\phi^M\phi^M\phi^I + \phi^M\phi^I\phi^M
+ \phi^I\phi^M\phi^M)\,.\cr}}
Using the notation of \refs{\Beisertone, \Beiserttwo},
we write the Hamiltonian as
\eqn\bozo{\eqalign{ H &= \sum_{i=1}^L 2 (\psi(J_{i,i+1} +1) - \psi(1)) 
\equiv\sum_{i=1}^L 2 \, h(J_{i, i+1}) \cr
&=\sum_{i=1}^L \sum_{j=0}^\infty 2 \,h(j) P_{i,i+1,\, j}\cr }}
where $P_{i,i+1,\, j}$ projects the fields at positions $i,i+1$ to
the module $V_j$. These modules, for $j=0,1,2,...$ label the
irreducible representations of $PSU(2,2|4)$ which describe the
two-particle system in free $N=4$ super Yang-Mills theory. 
They appear in the tensor product $V_F\otimes V_F = \sum_{j=0}^\infty V_j$,
as discussed in \crow.
The $h(j) = \sum_{n=1}^j {1\over n}$ 
are harmonic numbers (with $h(0)\equiv 0$).

For $L=2$, we know $(J^A_1 + J^A_2)^2 |\lambda_2\rangle= 0$,
since  $|\lambda_2\rangle$ is the lowest component of the module
$V_j$ with $j=0$, {\it i.e.} $J_{12}|\lambda_2\rangle = 0$, where 
$(J^A_1 + J^A_2)^2 = J_{12} (J_{12} + 1)$. Hence, as is well known, 
\eqn\cozo{H |\lambda_2\rangle = 2 h(0) (P_{12,\,0} + P_{21,\,0}) 
|\lambda_2\rangle = 4 h(0)  |\lambda_2\rangle =  0\,.}
In fact $H$ annihilates all states in the module $V_0$, the
superconformal chiral primary irreducible representation for two sites,
since all such states are accessed from 
$|\lambda_2\rangle$ by generators of  
$PSU(2,2|4)$ which commute with $H$. Since $U$ is a three-site operator,
it does not act on $|\lambda_2\rangle$.

The $L=2$ states which are lowest weights of the modules $V_j$, with $j>0$,
are superconformal primaries but not {\it chiral} superconformal primaries. 
The $j>0$ states are not protected, and they will receive quantum corrections 
to their conformal dimension. 

Now we consider chiral primaries with more than two sites 
$|\lambda_L\rangle$.
Since the chiral primary state is symmetric and traceless in all indicies,
it follows that it is symmetric and traceless in each pair. States on a pair
of sites have definite $j$ as described by\crow\ from 
\refs{\Beisertone, \DNW}, and
since each pair is symmetric and traceless in the scalar fields then $j=0$:
\eqn\projag{P_{i,i+1, 0} |\lambda_L\rangle = 
|\lambda_L\rangle\,,}
and projections to all other values of $j$ vanish, 
\eqn\projag{P_{i,i+1, j} |\lambda_{L}\rangle = 0\,,\qquad j\ne 0\,.}
Then
\eqn\speedo{\eqalign{ H |\lambda_{L}\rangle &=
\sum_{i=1}^{L} \sum_{j=0}^\infty 2 \,h(j) P_{i,i+1,\, j} \, |\lambda_{L}
\rangle\cr
&= \sum_{i=1}^{L}  2 \,h(0) P_{i,i+1,\, 0} \, |\lambda_{L}\rangle = 0
\,,}}
since $h(0)=0$.
For {\it eg.}, for the three-site chiral primary state $|\lambda_3\rangle$.
we find that the one-loop anomalous dimension vanishes, since 
\eqn\peds{\eqalign{ H |\lambda_3\rangle &= 
\sum_{i=1}^{L} \sum_{j=0}^\infty 2 \,h(j) P_{i,i+1,\, j} \, |\lambda_3\rangle
\,,}} and
\eqn\newproj{P_{i,i+1,j} |\lambda_3\rangle = \delta_{j,0} |\lambda_3\rangle\,,}
for $1\le i\le 3$.
To see this explicitly,
we can rewrite the chiral primary \oscar\ 
and identify it with  $P_{12,\,0}|\lambda_3\rangle$: 
\eqn\appy{\eqalign{ |\lambda_3\rangle
&=(\phi^I\phi^J + \phi^J\phi^I - {1\over 3}\delta^{IJ}\phi^M\phi^M )\,\phi^K\cr
&\hskip6pt
+(\phi^I\phi^K + \phi^K\phi^I - {1\over 3}\delta^{IK}\phi^M\phi^M )\,\phi^J\cr
&\hskip6pt
+(\phi^J\phi^K + \phi^K\phi^J - {1\over 3}\delta^{JK}\phi^M\phi^M )\,\phi^I\cr
&\hskip6pt
-{1\over 4} \delta^{IJ} (\phi^M \phi^K + \phi^K\phi^M 
-{1\over 3}\delta^{KM}\phi^N\phi^N )\,\phi^M\cr
&\hskip6pt
-{1\over 4} \delta^{IK} (\phi^M \phi^J + \phi^J\phi^M 
-{1\over 3}\delta^{JM}\phi^N\phi^N )\,\phi^M\cr
&\hskip6pt
-{1\over 4} \delta^{JK} (\phi^M \phi^I + \phi^I\phi^M 
-{1\over 3}\delta^{MI}\phi^N\phi^N )\,\phi^M\cr
&=  P_{12,\,0}\,|\lambda_3\rangle\,,\cr}}
{\it i.e.} when we project the fields at positions $1,2$ to the module
$V_0$, we find that $P_{12,\,0} |\lambda_3\rangle = |\lambda_3\rangle$
which checks that $P_{12,\,j} |\lambda_3\rangle = 0$ for $j\ne 0$.
Since $|\lambda_3\rangle$ is totally symmetric, it follows that  
$P_{23,\,0} |\lambda_3\rangle = |\lambda_3\rangle$,
and  $P_{31,\,0} |\lambda_3\rangle = |\lambda_3\rangle$.
Then from \peds,
\eqn\pedsa{\eqalign{ H |\lambda_3\rangle &=
2 h(0) ( P_{12,\,0} + P_{23,\,0} +  P_{31,\,0}) |\lambda_3\rangle = 0\,.}}

The second Casimir $U$ also annihilates $|\lambda_{L}\rangle$, 
since $U$ is a sum of commutators that fan out to a sum of products of
the Hamiltonian densities $2 h(J_{i,i+1})$, 
each of which annihilates the chiral primary.
Again we look at this explicitly for the lowest component 
of the three-site chiral primary $|\lambda_3\rangle$,$\,$
where
\eqn\abo{\eqalign{ U |\lambda_3\rangle &= 
\sum_{i=1}^{L} [h(J_{i,i+1}), h(J_{i+1,i+2})]  \, |\lambda_3\rangle\cr
&=\sum_{i=1}^{L} \sum_{j=0}^\infty \sum_{j'=0}^\infty
 \,h(j) h(j') [P_{i,i+1,\, j}, P_{i+1,i+2,\, {j'}}\,] \, |\lambda_3\rangle
\cr &=  \sum_{j=0}^\infty (\, h(j) h(0) 
(P_{12,\,j} P_{23,\,0} - P_{23,{j}} P_{12,\,0} \cr
&\hskip70pt + 
P_{23,\,j} P_{31,\,0}\, - P_{31,{j}}
P_{23,\,0} \cr 
&\hskip70pt + 
P_{31,\,j} P_{12,\,0}\, - P_{12,{j}} P_{31,\,0} \,)\,)\, |\lambda_3\rangle
=0\,.\cr}}
The fact that $H$ and $U$ annihilate the lowest component states 
$|\lambda_{L}\rangle$,
extends to all states in the superconformal chiral primary modules,
since $[H, J^A] = [U, J^A] = 0$. 

We conjecture that all the Casimirs of the Yangian annihilate the
chiral primaries. This would explain why 
we see only the ordinary $PSU(2,2|4)$ symmetry in
the supergravity Lagrangian of the AdS/CFT dual theory. 

\newsec{Higher-Loop Corrections}

We have conjectured in \refs{\DNW} that the $PSU(2,2|4)$ 
Yangian generators can be defined to all orders in $g^2 N$ in the planar 
limit of the ${\cal N} = 4$ SYM, where the exact generators obey
\eqn\yangt{\eqalign{
& [\tilde J^A, \tilde J^B]= f^{AB}_{C}\tilde J^C\,,\qquad 
[\tilde J^A, \tilde Q^B]= f^{AB}_C \tilde Q^C\,,}}
together with the Serre relations \refs{\DRone,\Bernardone\cinn}. 
Here we make a few comments on how \yangt\ can be verified and 
used to probe higher-loop corrections. 

The second equation in \yangt\ expanded to one-loop is
\eqn\incoco{
[\delta J^A,Q^B]+[J^A,\delta Q^B]=f^{AB}_C \delta Q^C\,.}
For $J^A = D$, \incoco\ becomes
\eqn\incocot{
[\delta D ,Q^B]+[D,\delta Q^B]=\lambda^B \delta Q^B\,,}
where $\lambda^B$ is the bare conformal dimension of $J^B$, and
$\delta D$ is the one-loop planar (spin chain) Hamiltonian that we have
discussed before.
The structure constants are given by the algebra,
so they do not receive quantum corrections.
Since $[D,\delta Q^B]=\lambda^B \delta Q^B$, and we checked in \refs{\DNW} that
$[\delta D ,Q^B] = 0$ modulo the edge effects which
vanish for an infinite or periodic chain, this means that we have
verified \yangt\  through one-loop for all tree generators 
$Q^A$ and a particular one-loop generator, the anomalous dimension 
operator $\delta D \equiv D_2 = H$.

We can also consider the one-loop correction to the non-local 
Yangian generators $\delta Q^B$. 
For the special bilocal Yangian generator associated with
the dilatation index, 
$Q^B = Q^{(D)}$ in \incoco, such a correction must satisfy
\eqn\ly{[\delta J^A,Q^{(D)}]+[J^A,\delta Q^{(D)}]= - \lambda^A  
\delta Q^A\,.}
We expect $\delta Q^{(D)}$ to depend on three sites at a time, and 
note that adding the second local Hamiltonian $\bar U$ to $\delta Q^{(D)}$ 
respects the algebraic constraints. But $\delta Q^{(D)}$ 
will also have non-local contributions.
It could have a form similar to a non-local abelian Hamiltonian.

If we could compute the one-loop correction to the Yangian generator
$\delta Q^A \equiv Q_2^A$, 
we could use it to check that 
\eqn\ordera{[D_4, Q^B] + [D_2, Q_2^B] = 0\,,}
where $D_2, D_4$ are the one and two-loop corrections to the dilatation
operator. This constraint follows from
$[\tilde J^A, \tilde Q^B] = f_{ABC} \tilde Q^C$, which expanded
to second order, ${\cal O}(g^2 N)^2$ in the `t Hooft coupling, and 
neglecting odd powers of $g$, is 
\eqn\orderf{[J^A_4, Q^B] + [J^A, Q^B_4] + [J^A_2, Q^B_2] = f^{AB}_C Q^C_4\,.}
The ordinary symmetry analogue of \ordera\ , $[D_4, J^B] + [D_2, J_2^B] = 0\,,$
holds by inspection for the dilatation $J_2^{(D)} \equiv D_2$, 
since $[D, D_4] = 0$.

Next we comment on the possibility that the
exact anomalous dilatation operator is an integrable Hamiltonian.
Using the conjecture \yangt, we will show that
the anomalous dimension $\Delta D$ to all orders in $g^2N$ is a Casimir
of the exact Yangian algebra.
We consider the entire anomalous dilatation operator
\eqn\ean{\Delta D = \tilde D - D\,,}
and note that since
\eqn\nolo{\eqalign{[D, Q^B ] = \lambda^B  Q^B\,,\qquad
[\tilde D, \tilde Q^B ] = \lambda^B \tilde Q^B\,,\cr}}
and higher-loop corrections retain their bare
conformal dimension,
\eqn\rolo{[D, \tilde Q^B ] = \lambda^B  \tilde Q^B\,,}
we find that
\eqn\polo{[\Delta D, \tilde Q^B] = 0\,.}
This argument parallels one in \refs{\Beisertone}
using the exact commutation relation for the ordinary symmetry generators,
resulting in
\eqn\polo{[\Delta D, \tilde J^B] = 0\,.}
Thus we find from the Yangian defining relations \yangt, that
a Casimir of the exact Yangian is given by $\Delta D$.
This strengthens the motivation to
identify $\Delta D$ with a higher-loop integrable Hamiltonian
Although the exact dilatation
operator $\tilde D$ does not commute with the exact Yangian being one
of its generators, the anomalous piece $\Delta D$ does.
This suggests that the exact anomalous dilatation operator is an
integrable Hamiltonian, which could be used to find the eigenvectors and
eigenvalues of the states in the exact superconformal gauge theory.

Having probed the equations defining higher-loop corrections
to the Yangian, we remark that
if we were able construct the monodromy matrix with a suitable
Yang-Baxter equation
in terms of the exact Yangian generators $\tilde J^A, \tilde Q^A$,
then the trace of the exact monodromy
$\tilde T_m (u, g^2N)$ could be reexpanded
in $u$ to find the family of Hamiltonians
at higher loops, $\tilde H_\kappa (g^2N)$.
In some subsectors,
commuting higher-loop higher Hamiltonians have been conjectured
as a ``best guess'' to define higher-loop integrability \refs{\Beisertfour,
\Beisertfive},
$[\tilde H_\kappa(g^2 N),  \tilde H_\rho(g^2 N)] = 0$.
This has been used to make an educated guess about the
three-loop anomalous dimensions of certain operators.
There is a possible discrepancy between this three-loop calculation and
the corresponding three-loop term in the AdS/CFT dual string theory
\refs{\Serban-\Zarembo}.
It would be interesting to find the connection between those higher-loop
Hamiltonians and the higher-loop Yangian
\refs{\Arutyunov-\Berktwo}. 

One might speculate that if we were to replace $J^A$ and $Q^A$
by the exact generators $\tilde J^A$, $\tilde Q^A$,
then \monoyang\ would become the two-site monodromy matrix for the
exact planar superconformal Yang-Mills theory, restricted to the
$SU(2)$ subsector, with a similar extension to $L>2$.
But for that speculation to hold,
$\tilde J^A$ and $\tilde Q^A$ would have to satisfy the condition \extra\ ,
as well as the Yangian defining relations. Possibly the extension
of the monodromy to higher loops is more complicated than this, and
involves additional $g^2N$ dependence beyond the simple dependence
through exact Yangian generators.

\newsec{Conclusions and Outlook}

Many of the features of integrable spin systems, such as 
a spin chain expression for the Hamiltonian, its commuting
family of local Hamiltonians, the non-abelian Yangian generators 
(including the ordinary symmetry generators), the R-matrix $R_{ij}(u)$,
the monodromy matrix $T_m(u)$, and the
trace of the monodromy matrix $F(u) = \tr_m T_{m}(u)$ 
can be constructed for the four-dimensional planar 
${\cal N} = 4$ superconformal $SU(N)$ gauge theory.
The techniques we use 
are novel in that they rely heavily on the Yangian.
Gauge invariant states in planar Yang-Mills theory correspond to
operators with a single trace. In their spin model description,
this is reflected by periodic boundary conditions and a zero momentum
condition. 
Although  our representation for the tree level
Yangian cannot be defined for periodic boundary conditions,
we found that these techniques are still useful
in the gauge theory. 

The way the explicit {\it spin chain}
expressions appear in the gauge theory is by 
the non-abelian non-local Yangian generators taking a tree level role, 
while the abelian local Hamiltonians are identified with one-loop expressions. 
In conventional spin models, 
the analytic properties of the monodromy matrix link these two 
sets of symmetries together \refs{\Bernardthree}. In the gauge theory,
since the interpretation of the structures now involve both tree and 
one-loop quantities, this monodromy intertwines the two lowest orders
of perturbation theory. 

The two sets of symmetries, abelian and non-abelian,
are also linked in the gauge theory by the identification of the periodic chain
one-loop dilatation operator $H$, and the second local abelian Hamiltonian $U$,
as Casimirs of the Yangian. 
The fact that $H$ is a Casimir was seen to follow from 
consistency conditions which arose in an expansion to one-loop,
when the ${\cal N}=4$ Yang-Mills theory
was assumed to have Yangian symmetry for all $g^2N$  \refs{\DNW}.
Together with the appearance of the non-abelian symmetry at $g^2 N = \infty$,
as indicated in the dual string theory\refs{\Bena,
\Polyakovone-\berk},
these results provide evidence for the presence of 
Yangian generators at higher loops. 
Their Casimir operators will then have higher-loop contributions,
and they will form a commuting set of periodic chain higher-loop Hamiltonians.
These will be local in the sense that the interactions involving
a certain number of sites will vanish
for sites far enough apart, since for periodic boundary conditions
that feature leads to 
a vanishing commutator with the Yangian generators. 
One of the exact Casimirs is the exact anomalous 
dimension operator $\Delta D$.

In conclusion,
the Yangian is an important tool, and computing its higher-loop corrections
may clarify the formulation of higher-loop integrability.
Of particular interest is the extension to large $g^2 N$ of the
Bethe ansatz equations for the eigenvalues and eigenvectors of 
 $\Delta D$.
It would also be valuable to develop the link between integrability
and twistor space structures in Yang-Mills theory \refs{\Witten-\Mason}.

\appendix {A} {\hskip3pt Action of $q_{12}^A$ on Two-Spin
States in the $XXX_s$ Spin Model} 

Two-spin states in the $XXX_s$ model form modules $V_j$ in the
tensor product decomposition 
\eqn\tp{V_S \otimes V_S = \bigoplus_{j=0}^{2s} V_j\,}
when $s=0,\half, 1,\ldots$. 
We define $q^A_{12} = J^A_1 - J^A_2$, and we will show
that $q^A_{12} V_j$ is contained in $V_{j+1} \oplus V_{j-1}$,
as found earlier for the $PSU(2,2|4)$ gauge theory \refs{\DNW}.
As before, we prove this in two parts: (1) 
$q_{12}^A V_j$ occurs in the direct sum of $V_k$ with $k-j$ odd,
and (2) $q_{12}^A V_j$ occurs in the direct sum of $V_k$ with
$|j-k|\le 1$.
Part (1) follows from the $SU(2)$ tensor product decomposition.
Let $\sigma$ be the operator that exchanges the two $V_S$ modules 
in \tp. 
The irreducible representations $V_j$ are either symmetric or antisymmetric,
and for a given value of $s$, $\sigma V_j = (-1)^{j+2s} V_j$.
Since $ \sigma q^A_{12} = - q^A_{12}$; and 
the action of $\sigma$ on  $q^A_{12} V_j$ must match that of $\sigma$ on $V_k$,
since $q^A_{12} V_j \in\oplus V_k$, we have
\eqn\dolly{\eqalign{\sigma\, 
q^A_{12} V_j = (-1)^{j+1+2s} q^A_{12} V_j \,,\cr
\sigma V_k = (-1)^{k+2s} V_k\,,\cr}}
which implies $(-1)^{j+1} = (-1)^k$ so $k-j$ must be odd. This proves
Part (1). 
For Part (2), it will be sufficient to consider only the highest weight state
$|\lambda(j)\rangle$ in each $V_j$, since any state in $V_j$ is related to
$|\lambda(j)\rangle$ by the raising operator $J^+$ of $SU(2)$, 
and the commmutator of $q^A$ with the raising operator is a linear combination
of the $q^A$. An arbitrary state in $SU(2)$ is labelled by $|j,m\rangle$,
where $m$ is the $J^3$ eigenvalue, and $j$ is the quadratic Casimir eigenvalue.
In this notation, the highest weight state is $|\lambda(j)\rangle = 
|j,j\rangle$,
with $J^+ |j,j\rangle  = 0$.
We will prove directly that 
\eqn\xpl{\eqalign{q^+  |j,j\rangle &\in V_{j+1}\,,\cr 
q^3 |j,j\rangle &\in V_j + V_{j+1}\,,\cr
q^- |j,j\rangle &\in V_{j-1} + V_j + V_{j+1}\,,\cr}}
which will verify Part (2). 
We first consider $q^+ |j,j\rangle$ in \xpl. Since $J^3 q^+ |j,j\rangle =
[J^3, q^+] |j,j\rangle + j q^+ |j,j\rangle = (j+1) |j,j\rangle$, 
then $q^+ |j,j\rangle \in \oplus_{k\ge j+1} V_k$, since modules with
$k\le j$ do not contain states with $m=j+1$ eigenvalues. 
Also, $J^+ q^+  |j,j\rangle = 0$, so $q^+ |j,j\rangle$ 
is a highest weight state,
thus $q^+  |j,j\rangle \in V_{j+1}$.
Similarly, $J^3 q^3  |j,j\rangle = j q^3  |j,j\rangle$, 
and $J^+ J^+ q^3 |j,j\rangle = 0$,
so $q^3 |j,j\rangle \in V_j + V_{j+1}$. 
Lastly, 
$J^3 q^- |j,j\rangle = (j-1) |j,j\rangle$ and $J^+ J^+ J^+ q^- |j,j\rangle = 0$,
so $q^- |j,j\rangle \in V_{j-1} + V_j + V_{j+1}$ which completes the proof of 
Part (2).

\vskip20pt
\noindent{\bf Acknowledgements:}

We thank Edward Witten for discussions.
CRN was partially supported by the NSF Grants PHY-0140311 and
PHY-0243680.
LD thanks Princeton University for its hospitality,
and was partially supported by the U.S. Department of Energy,
Grant No. DE-FG02-03ER41262.
Opinions and conclusions expressed here are those of the authors 
and do not necessarily reflect the views of the funding agencies.

\listrefs

\end